\documentclass[a4paper,11pt]{article}
\pdfoutput=1
\usepackage{jcappub}
\usepackage{bm}
\usepackage{soul}
\usepackage{latexsym}
\usepackage{dcolumn}
\usepackage{amsfonts,amssymb,amsmath}
\usepackage{graphicx,epsfig}
\usepackage{psfrag}
\usepackage{subfigure}
\usepackage{rotating}
\usepackage{hyperref}
\usepackage{tikz}

\hypersetup{
	unicode=false,          
	pdftoolbar=true,        
	pdfmenubar=true,        
	pdffitwindow=false,     
	pdfstartview={FitH},    
	pdftitle={My title},    
	pdfauthor={Author},     
	pdfsubject={Subject},   
	pdfcreator={Creator},   
	pdfproducer={Producer}, 
	pdfkeywords={keyword1} {key2} {key3}, 
	pdfnewwindow=true,      
	colorlinks=true,       
	linkcolor=red,          
	citecolor=cyan,        
	filecolor=magenta,      
	urlcolor=blue,           
	linktocpage=true
}

\newcommand{\beq}{\begin{equation}}
\newcommand{\eeq}{\end{equation}}
\newcommand{\bal}{\begin{aligned}}
	\newcommand{\eal}{\end{aligned}}
\newcommand{\bea}{\begin{eqnarray}}
\newcommand{\eea}{\end{eqnarray}}

\def\beq{\begin{equation}}
\def\efq{\end{equation}}
\def\br{\begin{eqnarray}}
\def\er{\end{eqnarray}}
\def\benu{\begin{enumerate}}
	\def\efnu{\end{enumerate}}
\def\nn{\nonumber}

\def\l{\left}
\def\r{\right}
\def\cR{{\cal R}}

\def\d{{\rm d}}
\def\i{{\rm i}}

\def\f{\frac}
\def\nn{\nonumber} 

\def\cR{{\mathcal{R}}}
\def\cS{{\mathcal{S}}}

\renewcommand{\arraystretch}{2}

\def\d{{\rm d}}

\renewcommand{\arraystretch}{1.2}

\def\high{\vphantom{\Biggl(}\displaystyle}

\newcommand{\be}{\begin{equation}}
\newcommand{\barr}{\begin{array}}
\newcommand{\earr}{\end{array}}
\def\bal#1\eal{\begin{align}#1\end{align}}

\makeatletter
\gdef\@fpheader{}
\makeatother

\begin{document}

\title{Comparing multi-field primordial feature models with the Planck data}

\author[1,2]{Matteo~Braglia,}
\author[3]{Xingang Chen,}
\author[4,5]{Dhiraj Kumar Hazra}

	\affiliation[1]{DIFA, Dipartimento di Fisica e Astronomia,
		Alma Mater Studiorum, Universit\`a degli Studi di Bologna,
		Via Gobetti, 93/2, I-40129 Bologna, Italy}

	\affiliation[2]{INFN, Sezione di Bologna,
		via Irnerio 46, I-40126 Bologna, Italy}
	
\affiliation[3]{Institute for Theory and Computation, Harvard-Smithsonian Center for Astrophysics, 60
Garden Street, Cambridge, MA 02138, USA}
\affiliation[4]{The  Institute  of  Mathematical  Sciences,  HBNI,  CIT  Campus, Chennai  600113,  India}
   \affiliation[5]{INAF/OAS Bologna, Osservatorio di Astrofisica e Scienza dello Spazio,
		Area della ricerca CNR-INAF, via Gobetti 101, I-40129 Bologna, Italy}

\emailAdd{matteo.braglia2@unibo.it, xingang.chen@cfa.harvard.edu, dhiraj@imsc.res.in}

	\abstract
{In this paper, we use a complete model of classical primordial standard clocks as an example to develop a methodology of directly comparing numerical predictions from complicated multi-field feature models with the Planck data, including the Planck 2018 Plik unbinned likelihood and the statistically most powerful CamSpec 2020 likelihood for temperature and polarization data. As this two-field inflationary model offers a plethora of primordial feature spectra that represent combinations of sharp and resonant feature signals non-trivially distributed over extended cosmological scales, its data comparison has not been satisfactorily addressed by previous attempts using analytical templates. The method of this paper, consisting of numerical prediction, effective parameter construction and nested sampling data comparison, allows us to efficiently explore every possible spectra from the model. We classify the resulting feature candidates in three different frequency ranges. We use the Bayesian evidences to assess the statistical significance of the candidates over the baseline model, taking into account the effect of additional parameters and the look-elsewhere effect.
Although none of the candidates is statistically significant, the methodology of this paper can be used to facilitate the future model-building and data-screening process of primordial features, and the  candidates can be subjected to further tests with data from the upcoming cosmic microwave background polarization observations and galaxy surveys. }

	\maketitle

\section{Introduction}
\label{Sec:Introduction}
\setcounter{equation}{0}

Primordial features are an important class of observational signatures of the beyond-the-standard-model cosmology. Observationally, primordial features are small and scale-dependent departures from the leading order scale-invariant spectra of the primordial density perturbations. These departures are often oscillatory functions of scales. In model-building, these features are induced by features in the potential or internal space of primordial Universe models, which disturb the background evolution away from its leading order trajectory. For example, in inflation scenario, such leading order trajectories are the attractor solutions, and the disturbs away from the attractor solution may be temporary or persistent as long as the attractor trajectory remains as the leading order approximation. There are many reasons that features should be present generically in the primordial Universe, although the amplitudes of the signals are unknown. If observed, primordial features can provide crucial insights into the physics of the primordial Universe. See \cite{Chen:2010xka,Chluba:2015bqa,Slosar:2019gvt} for reviews.

It is not an easy task to search for signals of primordial features in data, especially in situations where the signal-to-noise ratios are low. 
Primordial features are important examples of models in which the model-dependent background evolution plays an important role in the predictions of the density perturbations.
When broadly classified into several categories, the features in each category indeed share some common characters which may be used to construct simple templates and it is a useful approach to search for these templates in a model-independent fashion \cite{Chen:2010xka}. However, if the signal-to-noise ratio is low or more details of the feature become important, explicit model building becomes necessary. For example, while all features from the category of sharp feature models share the common character of sinusoidal running in the momentum space, the envelop of this running behavior is very model-dependent. It depends on the detailed nature and configuration of the sharp feature in each model (for examples, see \cite{Starobinsky:1992ts,Adams:2001vc,Chen:2006xjb,Bean:2008na,Achucarro:2010da, Hazra:2010ve,Adshead:2011jq,Gao:2012uq,Miranda:2012rm, Hazra:2014goa,Fumagalli:2020nvq,Braglia:2020taf}). In general, numerical computation is necessary to generate accurate predictions. For models in the category of classical primordial standard clocks \cite{Chen:2011zf,Chen:2011tu,Chen:2014joa,Chen:2014cwa}, the signals consist of both the type of sharp feature signals and that of resonant feature signals, distributed in the longer and shorter scales respectively, with a smooth transition in between. Analytical expressions are unavailable so far to even approximate the entire running behavior, not mentioning to accurately describe the highly model-dependent entire signal.

To compare theories with experiments, these examples emphasize the importance of developing the pipeline that directly connects the explicit model building in the Lagrangian, prediction making through numerical computations, and data comparison. In this paper, we would like to illustrate this approach by studying a two-field model of inflationary classical primordial standard clock.

During inflation, as the inflaton travels along a low energy effective path, from the point of view of the inflationary landscape whose structure is determined by some unknown high energy physics, it is natural to imagine that the inflaton trajectory may be very complicated. From time to time, the inflaton may be distracted from a period of attractor evolution by features present in the landscape, and then settle down in another attractor evolution. During these transitions and depending on the natures of these sharp features, various massive fields, whose presence is necessary to provide the low energy effective path for the inflaton in the first place, may be temporarily excited and oscillate for a few $e$-folds. If the imprints of these phenomena in the density perturbations are significant enough, we can hope to observe them.
These seemingly by-products of the realistic inflationary model-building have been shown to encode a very important piece of information -- the functional form of the time dependence of the scale factor of the primordial Universe, hence a direct evidence for the inflationary scenario \cite{Chen:2011zf,Chen:2011tu,Chen:2014joa,Chen:2014cwa}. This type of models, along with their cousins in the non-inflationary scenarios, are collectively named as the classical primordial standard clocks (CPSC).\footnote{Quantum fluctuations of these massive fields provide a different type of observational signatures that may be used for the same purpose of measuring the scale factor evolution \cite{Chen:2015lza,Chen:2018cgg}, which we do not study in this paper.}

Due to the aforementioned difficulties,  the comparison between the theoretical predictions of CPSC models and observational data has been limited to the following two incomplete steps \cite{Chen:2012ja,Chen:2011tu,Chen:2014joa}. In the first step, only part of the signal that is less model-dependent, namely the clock signal part, has been approximated by a template and compared with data. In the second, several parameters of a particular model are fixed to the best-fit values identified from the first approach, in such a way that the model becomes simple enough and an analytical template, that represents the prediction of this model in a significantly reduced parameter subspace, becomes available. While being able to provide some interesting candidates, these methods drastically limit the efficiency of the interactions between theories and observations. To complete the analyses and to prepare for the data from cosmic microwave background (CMB) polarization and large-scale structure, in this paper, we would like to test the method of directly connecting the full prediction of the multi-field feature model with data. While primordial feature models generically have correlated predictions in both power spectrum and non-Gaussianities, in this paper, we restrict our attention to the primordial power spectrum.

Comparing the model to data, however, is far from trivial. 
Firstly, full numerical evolution of the curvature and isocurvature perturbation equations is necessary to confront the full model with the data. We use two-field model extension of BINGO (BI-Spectra and non-Gaussianity Operator)~\cite{Hazra:2012yn} that was developed in~\cite{Braglia:2020fms}. The success of such a method will also greatly facilitate the process of constructing new models, being either CPSC or other multi-field feature models, that can be efficiently subject to experimental tests. 

The numerical implementation of the model Lagrangian in the code, which itself already presents complication, is not enough for an optimal data analysis. Indeed, our CPSC model is described by many parameters and some of which are strongly degenerate. In order to efficiently constrain the space of the feature candidates, two steps are necessary: i.e., the identification of the parameter combinations that have independent observational effects and a clever choice of priors. Furthermore, given the vast pool of candidates that the model is able to produce, multi-modality of the posterior distributions of the parameters is expected. Common sampling techniques used for cosmological parameter inference such as  Markov-Chain Monte Carlo are not suitable to the exploration of such a complicated prior volume and we resort to nested sampling.
 
It is needless to mention that, at the present stage, CMB data from Planck is the best data to constrain inflationary physics.   
The partial analytical templates of the CPSC model were compared with Planck 2013 data \cite{Chen:2014joa,Chen:2014cwa}. While Planck 2013 release was limited to temperature anisotropy data, in 2015 and 2018 both temperature and polarization data were released in public, with 2018 including the large scale polarization data from Planck high frequency instrument (HFI). Polarization data and its correlation with temperature data provide different perspectives. With the complete datasets now we can test the significance of primordial feature candidates that were only supported by temperature data. Features that are supported by multiple datasets have the best chances to represent {\it physical} features while we can rule out other features that do not improve the fit to the complete likelihood compared to the baseline model. The Planck 2018 official release~\cite{Aghanim:2019ame} included the {\tt Plik} and {\tt CamSpec} likelihoods and the unbinned version of  {\tt Plik} was used for the inflationary feature analysis by the Planck team~\cite{Akrami:2018odb}. Moreover, after the Planck release, the {\tt CamSpec 12.5HMcln}~\cite{Efstathiou:2019mdh} likelihood was released that provided statistically more powerful likelihood compared with earlier releases with 80\% coverage in both temperature and polarization sky. These two likelihoods present the opportunity to test the feature candidates to the same observations using two different statistical platforms with separate masking and cleaning procedures.

This paper is organized as follows. 
We start out in Sec.~\ref{Sec:CPSC} by reviewing our two-field feature model example -- a CPSC model originally introduced in Refs.~\cite{Chen:2014joa,Chen:2014cwa}. Our strategy to tackle the comparison of the model to the data is outlined in the subsequent several subsections of Sec.~\ref{Sec:Method}. In particular, in Sec.~\ref{subsec:Numerical} we detail the numerical integration used to arrive at the primordial power spectrum; in Sec.~\ref{subsec:effective_parameters} we identify the effective parameters governing different observational characteristics of the features; and in Sec.~\ref{subsec:DataAnalysis} we describe the nested sampling technique, the datasets and the priors on the parameters, and propose three different methods to find the bestfit feature candidates. These techniques are applied to the comparison against CMB data in Sec.~\ref{Sec:ComparePlanck}, where we show the statistics of our results and present our bestfit candidates for different ranges of frequencies of the clock signal.
We conclude in Sec.~\ref{Sec:Conclusion}.

To provide a roadmap for readers, we emphasize that there are two purposes in this paper. The first is to experiment with a methodological pipeline which can be used to screen complicated feature models for future purpose. The second is to constrain such a multifield feature model example. Readers who are only interested in the phenomenological aspects and final constraints on the model example used in this paper, may only read Sec.~\ref{Sec:CPSC} and~\ref{Sec:Pert} of the Sec.~\ref{Sec:Method}, and then directly proceed to the results presented in Sec.~\ref{Sec:ComparePlanck}.

\section{Methodology}
\label{Sec:Method}
\setcounter{equation}{0}

\subsection{A CPSC model and background evolution}
\label{Sec:CPSC}

A classical primordial standard clock model consists of at least two stages of inflation within the observable scales. These two periods of inflation are connected by a sharp feature which excites one or more massive fields out of the equilibrium. These massive fields start to oscillate classically and settle down again after a few $e$-folds. Note that, during this transition phase, the Universe continues to inflate and the effect of the sharp feature and massive fields is only perturbative. As introduced in Sec.~\ref{Sec:Introduction}, it is the imprints of these massive field oscillations, along with the effects of the sharp feature, on the density perturbations that interest us the most.

There are many possible ways to realize the above general picture. The resulting signals in terms of primordial features in the density perturbations have both model-dependent and independent properties. The full signal is always a combination of a sharp feature signal and a clock signal, with the former distributed in the larger scales and the latter in the shorter scales. The schematic running of each signal is also model-independent, with the former being the sinusoidal running and the latter the inflationary resonant running. 
On the other hand, the smooth connection between the sharp feature signal and the clock signal, the envelop of the sinusoidal running, as well as the relative amplitudes between the two signals, are very model-dependent. 
Although all of these belong to beyond-the-standard-model signatures, the clock signal is the most interesting part because it encodes the time dependence of the scale factor of the Universe. 
(See Ref.~\cite{Chen:2014cwa} for a more detailed summary.)
However, in practice, when the signal-to-noise ratio is low or because we would like to understand the full physical process, a comparison between full predictions of the model and data is desirable.

In this paper, to concretely explore the methodology, we use a two-field example of CPSC models, first introduced in Refs.~\cite{Chen:2014joa,Chen:2014cwa}. In this model, the sharp feature is represented by the inflaton falling into a potential well and starting to oscillate in a massive field direction while rolling in the inflaton direction of the second stage of the inflation. Because the curvature at the bottom of the potential is larger than the Hubble rate, the massive field, before settling down, oscillates for a few $e$-folds, possibly  generating a resonant feature signal in the primordial power spectrum.

The Lagrangian of the model is the following~\cite{Chen:2014joa,Chen:2014cwa}
\begin{equation}
	\mathcal{L}=-\frac{1}{2}\left(\tilde{R}+\sigma\right)^2 g^{\mu\nu}\partial_\mu\theta\partial_\nu\theta-V_{\rm inf}+\frac{m_\theta^2}{2}\tilde{R}^2\theta^2-\frac{g^{\mu\nu}}{2}\partial_\mu\sigma\partial_\nu\sigma-V_{\sigma 0}\left[1-\exp\left(-\sigma^2/\sigma_f^2\right)\right].
	\label{eq:LagrOrig}
\end{equation}

The role of the massive field here is played by $\sigma$, which we initially put on the plateau of its potential. We require that the flatness of the plateau  is such that it can only support the first $e$-folds of observable inflation and $\sigma$ eventually oscillates around its minimum during the subsequent $e$-folds, both of which can be tested by CMB observations. As $\sigma$ settles down in its minimum at $\sigma=0$, the trajectory approaches an effective single field one driven by the lighter field $\theta$. Note that, although the model in Eq.~\eqref{eq:LagrOrig} looks similar to a two-field example of quasi-single field inflation models studied in Ref.~\cite{Chen:2009we,Chen:2009zp}, here the mass of  $\sigma$  is much larger than  $H$ and, as just explained, our model contains an early phase that describes how the inflaton settles down into the bottom of its potential.

				\begin{figure}
		\begin{center} 
			\resizebox{214pt}{172pt}{\includegraphics{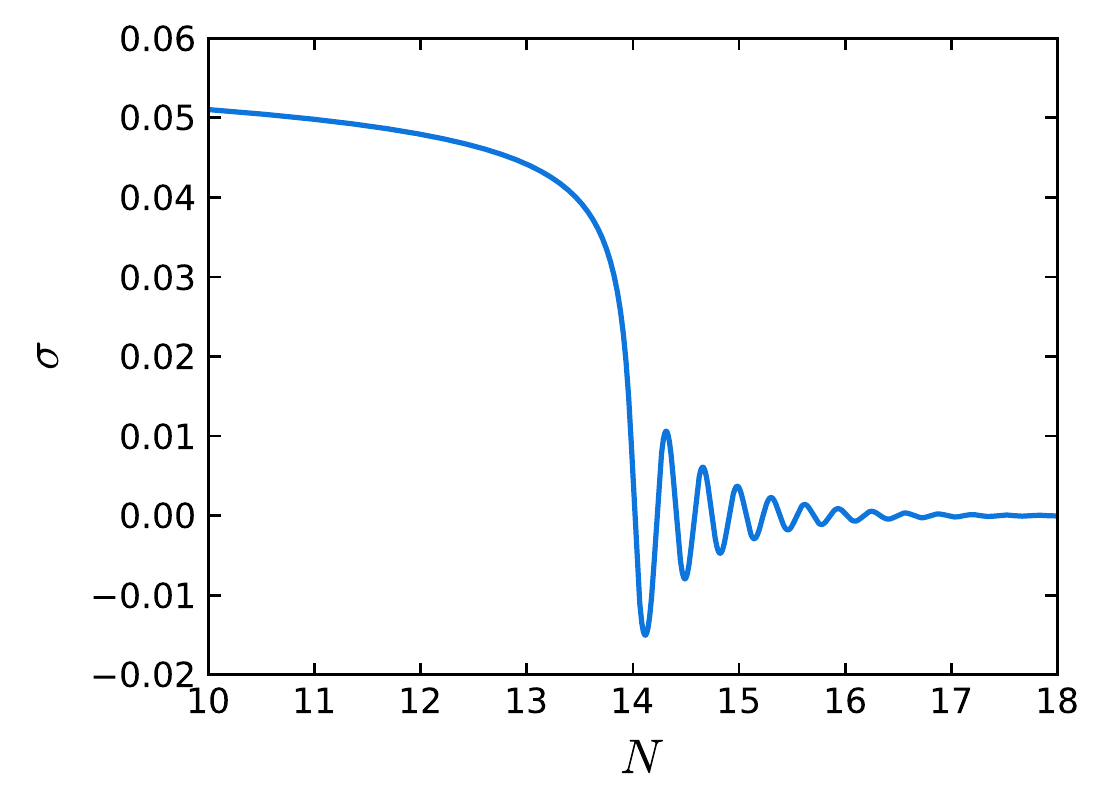}}
			\resizebox{214pt}{172pt}{\includegraphics{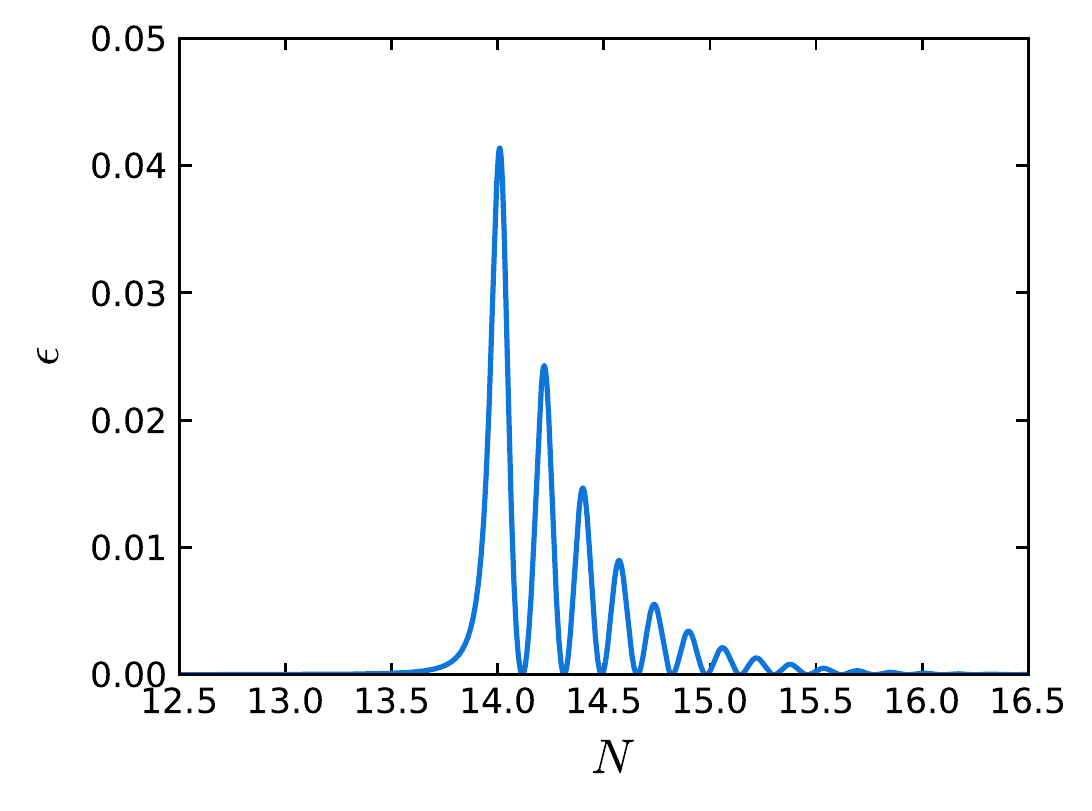}}	
		\end{center}
		\caption{\label{fig:backgroundIntro} 
		Example of the evolution of the clock field  [left] and the first slow-roll parameter $\epsilon$ [right] for our model. The parameters used in this plot are $V_{\rm inf}=4.73\times10^{-14},\,C_\sigma=0.02,\,\Theta_i=0.024,\,m_\sigma/H=20,\,C_\Theta=0.0185,\,\xi\sigma_f=0.003,\,\sigma_i=5.6\times10^{-2}$. }
	\end{figure}
	
	For later convenience, we find it useful to slightly rearrange the Lagrangian by canonically normalizing the field $\theta$ as
\begin{equation}
\Theta\equiv\tilde{R}\theta ~,
\end{equation}
and adsorbing $\tilde{R}$ in the kinetic coupling
\begin{equation}
\xi\equiv\frac{1}{\tilde{R}} ~.
\end{equation}
Defining also
\begin{equation}
C_\Theta\equiv\frac{m_\theta^2}{V_{\rm inf}}~,\,\,\,\,\,\,\,\,\,\,\,\,\,\,C_\sigma\equiv\frac{V_{\sigma 0}}{V_{\rm inf}}~, \,\,\,\,\,\,\,\,\,\,\,\,\,\,
\end{equation}
the Lagrangian thus takes the following form:
\begin{equation}
\mathcal{L}=-\frac{1}{2}\left(1+\xi\sigma\right)^2 (\partial\Theta)^2-\frac{(\partial\sigma)^2}{2}+V_{\rm inf}\left\{\left(1- \frac{C_\Theta}{2} \Theta^2\right)-C_{\sigma }\left[1-\exp\left(-\sigma^2/\sigma_f^2\right)\right]\right\},
\label{eq:LagrNew}
\end{equation}
from which we can derive the background equations of motion:
	\begin{align}~\label{eq:KGphi}
&\ddot{\sigma}+3 H\dot{\sigma}+V_\sigma
=\xi(1+\xi\sigma)\dot{\Theta}^2,\nonumber\\
&\ddot{\Theta}+\left(3 H+ 2 \frac{\xi}{1+\xi\sigma} \dot{\sigma}\right) \dot{\Theta}
+\frac{1}{1+\xi\sigma}V_\Theta=0
\end{align}
\begin{equation}
    3  H^2=\frac{1}{2}(1+\xi\sigma)^2\dot{\Theta}^2+\frac{1}{2}\dot{\sigma}^2+V_{\rm inf}\left\{1-\frac{C_\Theta}{2}\Theta^2+C_\sigma\left[1-\exp\left(-\sigma^2/\sigma_f^2\right)\right] \right\}
\end{equation}
Here the time derivative is denoted by overdots.

We give an example of the evolution of the massive field $\sigma$ and of the first slow-roll parameter	$\epsilon=-{\dot{H}}/{H^2}$ in Fig.~\ref{fig:backgroundIntro}. The plots show that the oscillatory pattern of the field $\sigma$ leads to a transient deviation of $\epsilon$ from its slow roll attractor value.

\subsection{Perturbation theory}
\label{Sec:Pert}

The perturbation theory for multi-field inflation models is well developed. While for single field models it is sufficient to solve for the curvature perturbations, for two-field models, the coupling between the curvature and isocurvature perturbation becomes important \cite{DiMarco:2005nq,Byrnes:2006fr,Lalak:2007vi}. 
At the linear level (which is what we are interested in this paper), this coupling can be conveniently handled in terms of either equations of motion or the in-in formalism \cite{Chen:2015dga}. Numerically, because differential equations are more convenient than the integrals, we will use the equation-of-motion approach.

There are several equivalent choices of the gauge and intermediate variables. In this paper, we work in the Newtonian gauge, where, in the absence of anisotropic stresses as is the case for minimally-coupled scalar fields, the metric takes the following form:
    	\begin{equation}
	\d s^2=-(1+2\Phi)\,\d t^2+a^2(t)\, (1-2\Phi)\,\d{\bm x}^2 \,,
	\end{equation}
	where $ \Phi$ is the Bardeen potential and we consider only scalar perturbations to the metric.
We define linear combinations of the field perturbations $\delta\sigma$ and $\delta\Theta$ that represent the tangent and perpendicular perturbations to the field space trajectory, namely $\delta\Sigma$ and $\delta s$ as\footnote{Note that in the literature $\delta\sigma$ and $\theta$ are used to denote the perturbation tangent to the trajectory in the field space and the turning in the field space. In order not to create confusion with the fields in our Lagrangian, we use $\delta \Sigma$ and $\vartheta$ instead.} \cite{Gordon:2000hv,DiMarco:2002eb}, 
\begin{eqnarray}~\label{perts}
		\delta\Sigma&=&\cos\vartheta\, \delta\sigma
		+\sin\vartheta e^b\, \delta\Theta,\nn\\
		\delta s&=&-\sin\vartheta\,\delta\sigma
		+\cos\vartheta e^b\, \delta\Theta,
		\end{eqnarray}
where $\sin\vartheta=e^b\dot{\Theta}/\dot{\Sigma}$, $\cos\vartheta=\dot{\sigma}/\dot{\Sigma}$,  $\dot{\Sigma}^2=\dot{\sigma}^2+e^{2b}\dot{\Theta}^2$ and, to make contact with the literature, we have defined $e^b\equiv (1+\xi\sigma)$. 
Using these quantities, we define the curvature and isocurvature perturbations as ${\cal R}= \Phi+(H/\dot{\Sigma})\delta\Sigma$ and ${\cal S}=(H/\dot{\Sigma})\delta s$ respectively. 

Note that $\mathcal{R}$ is the gauge-invariant curvature perturbation in the comoving gauge, which is equivalent to the Bardeen curvature perturbation $\zeta$ on scales much larger than the Hubble radius \cite{Bardeen:1980kt}, so we will use $\mathcal{R}$ and  $\zeta$ interchangeably. 
Also note that, because this model eventually settles down to a single field inflation model, only the perturbations along the $\Theta$-direction survive. This means that there can be other choices of the intermediate variables besides \eqref{perts}. For example, in Refs.~\cite{Chen:2014joa,Chen:2014cwa}, besides using a different gauge, the intermediate variables are chosen to be always along the $\delta\sigma$ and $\delta\Theta$ directions. These two approaches give the same results, as expected.

For each mode, we solve directly the evolution equations for $\mathcal{R}$ and $\mathcal{S}$ \cite{Braglia:2020fms}, 
		\begin{eqnarray}\label{eq:RS}
		\ddot{\mathcal{R}}
		&+&\left(H+2\f{\dot{z}}{z}\r)\dot{\mathcal{R}}
		+\frac{k^2 }{a^2}\mathcal{R}
		=-\frac{2 V_s}{\dot{\Sigma}}\dot{\mathcal{S}}
		-\,2\biggl(-{\rm e}^{-b}  b_{\sigma }\cos^2\vartheta\, V_{\theta }
		+ \sin\vartheta b_{\sigma } V_{\Sigma}
		+\, V_{\Sigma s}+\f{\dot{\Sigma}}{H }V_s\biggr)\mathcal{S},\notag\\
		\ddot{\mathcal{S}}
		&+&\l(H+2\f{\dot{z}}{z}\right)\dot{\mathcal{S}}
		+\biggl\{\frac{k^2}{a^2}-2 H^2-\dot{H}+\frac{H \dot{z}}{z}
		+\frac{\ddot{z}}{z}-\dot{\Sigma}^2-\dot{\Sigma}^2 b_{\sigma }^2 \cos^2\vartheta
		-\dot{\Sigma}^2 b_{\sigma \sigma }+V_{ss}\nn\\
		&+&\,b_{\sigma } \l[4 \sin \vartheta \, V_s
		+(1+\sin^2\vartheta) V_{\sigma }\r]
		\biggr\}\cS
		=\f{2  V_s}{\dot{\Sigma}}\dot{\mathcal{R}},
		\end{eqnarray}
		where $z\equiv a\dot{\Sigma}/H$.
As customary in the literature, we impose initial conditions on the gauge-invariant Mukhanov-Sasaki variables $Q_\Sigma=(\dot{\Sigma}/H)\cR$ and $\delta s=(\dot{\Sigma}/H)\cS$. When the modes are deep inside the Hubble radius, except for cases with specific strongly coupled or correlated initial states \cite{Cremonini:2010ua,Chen:2015dga} that are not of our concern, $Q_\Sigma$ and $\delta s$ are decoupled at least initially and can be approximated using Bunch-Davies initial condition as,
	\begin{equation}
	\label{eq:ic}
	Q_\Sigma(\tau)
	\simeq \delta s(\tau)
	\simeq \f{1}{a(\tau)}\,\f{{\rm e}^{-\imath k \tau}}{\sqrt{2 k}},
	\end{equation}
where $\tau$ represents the conformal time. Using this initial condition we solve Eqs.~\eqref{eq:RS} from sub-Hubble scales ($k\gg aH$). To be consistent with the first-principle in-in formalism at the linear level \cite{Chen:2015dga}, we follow the procedure of Refs.~\cite{Tsujikawa:2002qx,Lalak:2007vi} and use two sets of initial conditions. In the first case, we impose Bunch-Davies initial condition for the curvature perturbation and assume no initial isocurvature perturbations and we denote the solutions as ($\cR_1,\cS_1$). In the second case we impose the opposite and denote the solutions as ($\cR_2,\cS_2$). The dimensionless power spectra for the curvature perturbation 
at $N=N_\mathrm{end}$ is defined as, 
	\begin{eqnarray}\label{eq:PR}
	\mathcal{P}_{\cR}(k)
	&=&\frac{k^3}{2\pi^2}
	\l(\lvert\cR_1\rvert^2+\lvert\cR_2\rvert^2\r)
	=\mathcal{P}_{\cR_1}(k)+\mathcal{P}_{\cR_2}(k) ~.
	\end{eqnarray}
	Note that this procedure has been shown in Ref.~\cite{Chen:2015dga} to be equivalent to summing up all tree-level diagrams for the two-point function non-perturbatively in the in-in formalism.

\subsection{Numerical solution}
\label{subsec:Numerical}

We evaluate the primordial power spectra numerically in the CPSC model. We use the 2 field version of BINGO~\cite{Hazra:2012yn}~\footnote{This version, which was also used to produce the results of Refs.~\cite{Braglia:2020fms,Braglia:2020eai,Braglia:2020taf},  is not yet publicly available.} to solve the model. We first evaluate the background evolution of the fields by solving the Klein-Gordon equations \eqref{eq:KGphi}, with the number of $e$-folds used as the time variable for our numerical integration. We impose initial conditions on the clock field $\sigma$ and the inflation $\Theta$  and compute the initial field derivatives with respect to time assuming slow-roll condition. Note that here a fast roll initial condition will not alter our analysis as the effects will be beyond the largest relevant cosmological scale. We solve the backgrounds with adaptive stepsize control. As discussed in~\cite{Braglia:2020taf}, here too we assume the end of inflation is achieved by cutting off the $\Theta$ potential with an hybrid field. We obtain the initial value of the scale factor (at $N=0$) by the condition that the pivot mode $k=0.05~\mathrm{Mpc^{-1}}$ leaves the Hubble radius 50  $e$-folds before the end of inflation. Apart from the determination of initial scale factor, the end of inflation is not used for the evaluation. $N_\mathrm{end}$ therefore just changes the location of the features and since we allow the initial conditions on the fields to vary (see next subsection), it is safe to assume a fixed $N_\mathrm{end}$, which we take to be $N_\mathrm{end}=68$ for definiteness.

While in principle we should integrate the equations till the end of inflation, in the CPSC model the isocurvature perturbation decays and stops sourcing the curvature perturbation when the modes are well outside the Hubble radius. Therefore it is sufficient to evaluate the perturbations till the time when $k\ll aH$, which enables us to significantly speed up the comparison of the model to data. In particular, we evolve the perturbations from $k\ge100\,aH$ to $k=aH/10^5$. Here we should discuss an important point regarding the initial time to start the integration. For the modes that satisfy $k=100\,aH$ around the oscillatory phase certain sub-Hubble evolution gets imprinted in the power spectrum. Here we make sure that we start the integration of all the modes well before the onset of the oscillatory phase and therefore we need to identify the oscillatory feature region. During the background evaluation we determine the region of intermediate oscillatory phase. In the class of model we are considering, apart from the oscillatory phase, $\epsilon_1(N)$ grows with time. Therefore the end of the oscillatory phase ($N_\mathrm{OE}$) can be defined as the time when $\epsilon_1(N)$ grows as we {\it go back in time}. The beginning of the oscillatory phase ($N_\mathrm{OS}$) is defined when $\epsilon_1(N)$ becomes a few times larger than its initial value. Once $N_\mathrm{OS}$ is defined, we make sure that all modes are integrated from an {\it $e$-folds} $N(k\ge100\,aH)
<N_\mathrm{OS}$. The perturbation equations are also solved with adaptive stepsize control.     

The integrator BINGO is interfaced with {\tt CAMB}~\cite{Lewis:1999bs,Howlett:2012mh} to obtain the CMB angular power spectra. We do not consider the tensor power spectra as the values of the parameters we consider in this analysis generate negligible tensors (few orders of magnitude less than the Planck+BK15 bound, see discussion in the next subsection). In principle we should evaluate the power spectrum for each of the $k$-modes explicitly. However, given the number of equations in the two-field model, the integration of the modes from extremely deep inside the Hubble radius makes the computation of the power spectrum for few thousands modes time-consuming. Therefore we compute the power spectra for few hundreds of modes and interpolate with insignificant loss of precision. Firstly, we divide the entire inflationary dynamics into three stages, namely pre-oscillation phase ($N<N_\mathrm{OS}$), oscillation phase ($N_\mathrm{OS}<N<N_\mathrm{OE}$) and post-oscillation phase ($N>N_\mathrm{OE}$). Note that the entire cosmological scales probed by Planck data fall within these stages. Following this, we divide the scales by characterizing the modes that leave the Hubble radius in these three stages of inflation. Since during non-oscillatory phases we have nearly flat power spectra, we sample the scales with fewer (about 100) points. Evidently the modes that leave Hubble radius in the oscillatory stage need to be finely sampled compared to other stages. Here  we sample the points depending on the frequency of oscillations determined by the mass of the clock field $m_\sigma/H$. We make sure that we use at least 500 $k$-points for sampling. When the frequency of oscillations becomes larger, we proportionally increase the samples. The post-oscillations phase is also sampled proportionally to the frequency, but overall we use lesser samples compared to the oscillation phase. Combining these three stages we sample a total of about 1000 or more $k$-points (depending on the frequency). For these points we numerically compute the power spectra by solving the perturbation equations. For intermediate points we use cubic spline interpolation. Within {\tt CAMB} we compute the angular power spectra for all multipoles without interpolation irrespective of the frequency of the CPSC power spectrum.

\subsection{Model parameters and effective parameters}
\label{subsec:effective_parameters}

The next important step in this pipeline is to choose efficient parameters and parameter space to compute the numerical results and compare with data.
For complicated models, the original model parameters in the Lagrangian are sometimes not the most efficient and clear choice. It helps to examine some qualitative relations between parameters and model predictions, in order to 1) reduce the degeneracies in the {\em model parameters} in the Lagrangian and 2) to construct {\em effective parameters} from these model parameters that each has an observable effect that is as distinctive as possible. 
The details of this step is mostly a model-dependent art, but there are some general guidelines that we can follow, which we illustrate as follows using this CPSC model.

In the Lagrangian~\eqref{eq:LagrNew}, there are {\em seven} model parameters: five in the Lagrangian, $\xi$, $V_{\rm inf}$, $C_\Theta$, $C_\sigma$, $\sigma_f$, and two initial conditions, $\Theta_\i$, $\sigma_\i$ (we set the initial velocities of these two fields to the slow-roll attractor solution as mentioned previously).

On the other hand, the power spectrum with a full standard clock signal is expected to be characterized by {\em five}  or {\em six} effective parameters, each of which governs a specific property of the model prediction: the leading order featureless power spectrum can be described by a power-law characterized by the overall normalization $P_{\zeta *}$ and spectral index $n_s$, where the subscript $*$ denotes a pivot scale; the standard clock signal is a scale-dependent signal, so it has a location in the momentum space; the oscillation frequency of the clock signal is characterized by the mass of the massive field in unit of the Hubble parameter, $m_\sigma/H$; the overall amplitude of the oscillatory feature can be represented by its peak value.\footnote{For general feature models, the sharp feature in the Lagrangian has parameters characterizing its size and sharpness, besides its location, which give rise to the amplitude and extension of the sharp feature signal. However, in this special model \eqref{eq:LagrNew}, the sharp feature configuration is completely fixed by the configuration of the trough and so the properties of the sharp feature signal are related to those of the clock signal, so there are no extra parameters.} Another more subtle parameter describes the amplitude of the power spectrum at the largest scales, which we will discuss shortly.

We need to express these effective parameters in terms of model parameters and fix the degeneracies between the model parameters.
We separate the power spectrum as follows:
\begin{equation}
P_\zeta=P_{\zeta 0}\left(1+\frac{\Delta P_\zeta}{P_{\zeta 0}}\right) ~,
\end{equation}
where $P_{\zeta 0}$ is the leading order power-law spectrum, approximately described by $P_{\zeta *}$ and $n_s$,
and $\Delta P_\zeta/P_{\zeta 0}$ denotes the oscillatory correction introduced by features.

\begin{figure}
	\includegraphics[width=\columnwidth]{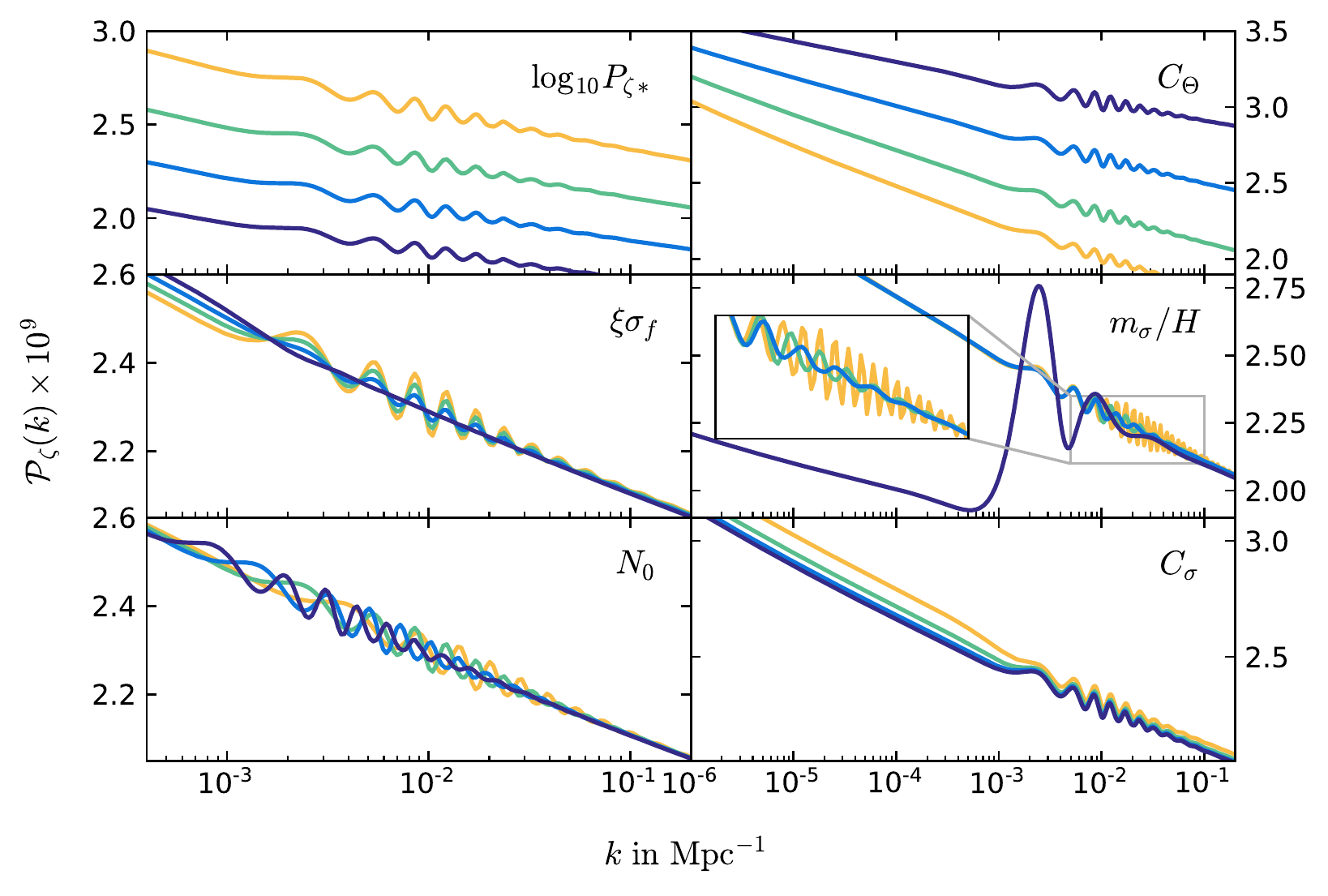}
	\caption{\footnotesize\label{fig:reparam} Effects of varying the effective parameters of the model on the primordial scalar power spectrum. We fix the baseline model to those of Fig.~\ref{fig:backgroundIntro} and vary one parameter at a time. Specifically, we vary  $\log_{10} P_{\zeta *}\in \{-8.5,\,-8.45,\,-8.4,\,-8.35\}$ in the top left-panel, $C_\Theta\in\{0.009,\,0.0135,\,0.0185,\,0.023\}$ in the top-right panel,  $\xi\sigma_f\in\{0,\,0.0015,\,0.003,\,0.0045\}$ in the center-left panel,  $m_\sigma/H\in\{5,\,10,\,30,\,50\}$ in the center-right panel, $N_0\in\{13,\,13.5,\,14,\,14.5\}$ in the bottom-left panel and   $C_\sigma\in\{0.005,\,0.01,\,0.02,\,0.04\}$  in the bottom-right one. In all the panels lighter colors correspond to increasing values of the parameters. In the center-right panel, for the $m_\sigma/H=5$ case, we use $\xi\sigma_f=0.07$ to highlight the possibility of producing Starobinsky-like features in our model.}
\end{figure}

The leading order part is as simple as the usual single field inflation model, because the inflaton $\Theta$ dominates the energy content of the Universe during inflation and at this order we can ignore the effect of the massive field and sharp feature. The model parameters $V_{\rm inf}$, $C_\Theta$ and $\Theta_i$ can be related to the following three observables:
\bea
P_{\zeta *} \approx \frac{V_{\rm inf}}{24\pi^2 \epsilon_\i}~,
~~~~~~~
n_s-1 \approx -6\epsilon_\i +2\eta_\i \approx -2 C_\Theta ~,
~~~~~~~
r \approx 16\epsilon_\i \approx 8 C_\Theta^2 \Theta_\i^2 ~,
\label{Eq:Pzeta_ns_expressions}
\eea
where $\eta_\i$ is the second slow-roll parameter, the small field approximation $\epsilon_\i \ll \eta_\i$ has been used, and the subscript ``i'' indicates the initial time $t_\i$ and  ``$*$" denotes the mode crossing the horizon at $t_\i$.
The effect of the parameters $P_{\zeta*}$ and $C_\Theta$ are illustrated in the top-left and top-right panels in Fig.~\ref{fig:reparam} respectively.
Currently we only have an  upper bound on the tensor-to-scalar ratio $r$, and this gives rise to a degeneracy. On the other hand, we note that in this model as long as the field is small, $\epsilon <10^{-6}$, the overall scale of the inflationary potential $V_{\rm inf}$  (or equivalently $\epsilon$) has no observable effect on the power spectrum \cite{Chen:2014cwa} and it only sets the mass scale for the whole model. It is thus better to fix this degeneracy by assigning $r$ a fixed value. In this work, we will fix
\bea
\epsilon_i=10^{-7} ~.
\label{Eq:epsilon_value}
\eea
Through \eqref{Eq:Pzeta_ns_expressions} and \eqref{Eq:epsilon_value}, the model parameters $V_{\rm inf}$, $C_\Theta$, $\Theta_i$ are related to the effective parameters $P_\zeta$ and $n_s$.

We next look at the parameters associated with the oscillatory feature in this model. 
The frequency of the background clock oscillation is given by the effective mass of $\sigma$, which, in unit of the Hubble parameter, can be approximated as
\begin{equation}
\label{eq:msigma}
\frac{m_\sigma}{H}\equiv\frac{V_{\sigma\sigma}(\sigma=0)}{H}=\frac{1}{H}\frac{\sqrt{2 C_\sigma V_{\rm inf}}}{\sigma_f}\approx\frac{\sqrt{3}}{\sqrt{V_{\rm inf}}}\frac{\sqrt{2 C_\sigma V_{\rm inf}}}{\sigma_f}=\frac{\sqrt{6 C_\sigma}}{\sigma_f} ~.
\end{equation}
The dependence of the clock signal on $m_\sigma/H$ is shown in the center-right panel of Fig.~\ref{fig:reparam}. Note that when $m_\sigma/H$ is very small, $\sigma$ undergoes only a few strongly damped oscillations around its minimum and we obtain a feature similar to that in single field models where the potential of the scalar field has a discontinuity \cite{Starobinsky:1992ts}.

The peak amplitude of the clock signal can be estimated using the in-in formalism and is approximately given by  \cite{Chen:2014cwa}
\begin{equation}
\label{eq:clockamplitude}
\frac{\Delta P_{\zeta}}{P_{\zeta 0}}\sim\sqrt{2\pi}\,\xi\sigma_f\,\sqrt{\frac{m_\sigma}{H}} ~.
\end{equation}
The proportionality of the amplitude to $\xi\sigma_f$ can be clearly seen from the center-left panel of Fig.~\ref{fig:reparam}.

The location of the feature in the momentum space can be specified by an effective parameter $N_0$, which is the number of $e$-folds since the initial time $t_\i$ till the inflaton encounters the sharp feature (i.e.~falls into the trough $\sigma=0$). $N_0$ can be related to the initial position $\sigma_\i$ by a shooting algorithm, in which we vary $\sigma_\i$ until the targeted value for $N_0$ is reached with a precision $\Delta N_0<0.005$.
The dependence of the location of the feature on $N_0$ is shown in the bottom-left panel of Fig.~\ref{fig:reparam}.
Equivalently, the location of the full clock signal can also be specified by the first mode of the sharp feature signal $k_0$ or the first mode of the resonant clock signal $k_r$ \cite{Chen:2011zf,Chen:2014cwa}. For convenience, we will also provide these values for the bestfit models.

The last parameter we would like to discuss is $C_\sigma$, the depth of the trough relative to the inflationary potential. From the qualitative analysis, Eq.~\eqref{eq:msigma} and \eqref{eq:clockamplitude}, we can see that fixing $m_\sigma/H$ and $\Delta P_\zeta/P_{\zeta 0}$ still leaves one degenerate parameter, which can be chosen as $C_\sigma$. More detailed numerical calculation shows that indeed this parameter only has some weak effect on the power spectrum at the largest scales and its amplitude, see bottom-right panel of Fig.~\ref{fig:reparam}. Because of this effect, we will still keep $C_\sigma$ as a variable, but we will see a large degeneracy associated with this parameter.

To summarize the above discussion, we conclude that it is convenient to impose priors on the following six effective parameters (fixing $\epsilon_\i=10^{-7}$): $\{ P_{\zeta *}, C_\Theta, C_\sigma, m_\sigma/H, \xi\sigma_f, N_0 \}$, which have been identified to each specify a property of the model prediction. Once values of these parameters are given, we use the following relations to determine the input values of the model parameters used in the MCMC analyses (note that $C_\Theta$ and $C_\sigma$ are already model parameters):
\bal
  V_{\rm inf} &= 24\pi^2 \epsilon_* P_{\zeta *} ~, \\
  \Theta_\i &= \high{ \frac{\sqrt{2\epsilon_\i}}{C_\Theta} } ~, \\
  \sigma_f &=\high{ \frac{\sqrt{6 C_\sigma}}{m_\sigma/H}}  ~, \\
  \xi &=\high{ \frac{\xi \sigma_f}{\sigma_f} } ~, \\
  \sigma_\i: & {\rm ~from~shooting~on}~N_0 ~.
\eal
The number of effective parameters is six, which is four more than that of the baseline (power law $\Lambda$CDM) model.

\subsection{Method of data analysis}
\label{subsec:DataAnalysis}

We use publicly available latest data of CMB temperature and polarization anisotropy from the Planck mission. Planck has mapped our Universe from the largest scales up to nearly $0.07^{\circ}$ that corresponds to multipoles $\ell=2-2500$. The Planck likelihood is divided between low-$\ell$ and high-$\ell$ both in temperature and polarization. For the low-$\ell$ TT (lowT), we use the $\tt{commander\_dx12\_v3\_2\_29}$ likelihood for ranges $\ell=2-29$. For low-$\ell$ EE (lowE), we use the HFI likelihood $\tt{simall\_100x143\_offlike5\_EE\_Aplanck\_B}$ spanning the same range. Since the CPSC model can produce high frequency oscillations in the power spectra, we do not use binned likelihood for high-$\ell$. For high-$\ell$ we use 2 different likelihoods for each of the TT, TE, EE and TTTTEEE likelihoods. The feature analysis of the Planck Inflation paper~\cite{Akrami:2018odb} used completely unbinned likelihood named {\tt Plik bin1} likelihood. Hereafter, we refer to this likelihood as P18 likelihood. P18 likelihood requires up to 21 (the number depends on the type of data being used) nuisance parameters. 100, 143 and 217 GHz auto and cross-spectra are used in the likelihoods. 

Following the Planck official release, the {\tt CamSpec} likelihood has been released in~\cite{Efstathiou:2019mdh}. The new of {\tt 2020 CamSpec release}\footnote{\href{https://people.ast.cam.ac.uk/~stg20/camspec/index.html}{https://people.ast.cam.ac.uk/~stg20/camspec}} provides the 12.5 version of CamSpec Planck likelihood which the authors claim to be the {\it ``most powerful"} likelihood. This likelihood makes use of $143\times143$, $143 \times 217$ and $217 \times 217$ TT spectra and utilizes 80\% of sky fraction using {\tt mask80}. Temperature spectra are cleaned with 545GHz maps and polarization spectra are cleaned with 353GHz maps. This likelihood does not use 100GHz auto spectrum and adopts a cleaning strategy which reduces the nuisance parameters substantially -- therefore, the TTTEEE likelihood contains only 9 nuisance parameters, 6 (3) of which are foreground (calibration) parameters. Hereafter we will denote this likelihood as EG20 likelihood. 

Given the qualitative differences between the P18 and EG20 likelihood, a feature analysis against both of them has the advantage of corroboration. Note that systematic effects present in one likelihood in the form of spurious oscillations may not be supported at similar statistical level, while potential feature candidates are expected to provide similar statistical fit to data in both likelihoods. In both the likelihoods the high-$\ell$ TT data are used within $\ell=30$ to $\ell\sim2500$ ($\ell\sim2508$ in P18) and TE and EE are used within $\ell=30$ to $\ell\sim2000$ ($\ell\sim1996$ in P18). We use 4 different data combinations, namely TT+lowT+lowE, TE+lowE, EE+lowE and TTTEEE+lowT+lowE, where the first terms in each data correspond to high-$\ell$ likelihood. We allow all the nuisance parameters to vary in our analyses apart from one particular method of best fit search that will be discussed in the subsequent sections.

\subsubsection{Nested sampling}
As we have discussed in Sec.~\ref{Sec:CPSC}, a plethora of primordial scalar spectra can be generated from the full CPSC model. Given this vast pool of candidates, it is expected that the posterior distribution of the parameters of CPSC model is multi-modal. Therefore instead of using Markov-Chain Monte Carlo, we use nested sampling through {\tt PolyChord}\footnote{\href{https://github.com/PolyChord/PolyChordLite}{https://github.com/PolyChord}}~\cite{Handley:2015fda,Handley:2015vkr}. Apart from estimation of parameters, the Bayesian evidence computed 
using {\tt PolyChord} helps in model comparison. The latter, also called marginalized likelihood, is the probability $Z\equiv p(\mathbf{x}|{\cal M})$ of observing the data $\mathbf{x}$ given the correctness of the model ${\cal M}$. From the ratio of the evidence of two separate models, it is possible to compute the Bayes factor as:
\begin{equation}
B_{12} \equiv\frac{Z_1}{Z_2}= \frac{\int d\boldsymbol{\theta}_1\, \pi(\boldsymbol{\theta}_1 \vert {\cal M}_1) {\cal L}(\mathbf{x} \vert \boldsymbol{\theta}_1,{\cal M}_1)\,}{\int d\boldsymbol{\theta}_2\, \pi(\boldsymbol{\theta}_2 \vert {\cal M}_2) {\cal L}(\mathbf{x} \vert \boldsymbol{\theta}_2,{\cal M}_2)\,}\,,
\label{eq:Bayes}
\end{equation}
where $\pi(\boldsymbol{\theta}_{\rm i})$ is the prior for the parameters ${\theta}_{\rm i}$ and ${\cal L}(\mathbf{x} \vert \boldsymbol{\theta}_{\rm i})$ the likelihood of the data given the model ${\cal M}_{\rm i}$. With our conventions, a positive $\ln$(Bayes factor) is interpreted as a Bayesian preference of the model  1 over the model 2. The extent of this preference can be read off from the so-called Jeffreys' scale, see Ref.~\cite{Kass:1995loi}.
In the following we will identify the CPSC model with 1 and the baseline $\Lambda$CDM model with 2. 

 The sampling works with $n_\mathrm{live}$ points. The algorithm updates these points in a way that shrinks the volume of this n-dimensional parameter space around the peaks of the posterior distribution. In our analyses we have used 1000 live points. From the samples generated by {\tt CosmoChord} ({\tt PolyChord} add on of {\tt CosmoMC}\footnote{\href{https://cosmologist.info/cosmomc/}{https://cosmologist.info/cosmomc/}} \cite{Lewis:2002ah}) we obtain the constraints on the parameters. Since the algorithm explores the boundary (corners) of the parameter space and compresses in each iteration, it is important to exclude the parameter spaces that are unphysical or expected to be well outside the posterior distribution. Therefore an optimized prior volume  is important. In Tab.~\ref{tab:priors} we provide the priors on the background parameters and the parameters in the CPSC model, while the priors on the nuisance parameters are kept at their default values. The priors on the background is kept in such a way that their posterior tails stay well within the ranges even in the datasets with least constraining ability, in this case EE+lowE. Since we are using lowE likelihood from HFI polarization maps, the $\tau$ prior we have used is much narrower compared to Planck 2015 constraints which did not use HFI low-$\ell$ polarization data. 
 
Following the discussions in Sec.~\ref{subsec:effective_parameters}, the priors on the effective parameters are chosen as follows: $P_{\zeta *}$ and $C_\Theta$ mainly control the amplitude and the tilt of the primordial power spectrum respectively, though $C_\Theta$ slightly changes also the amplitude along with the tilt which makes it somewhat degenerate with $P_{\zeta *}$. 
$P_{\zeta *}$ priors can simply be translated from the amplitude parameter in the power law $\Lambda$CDM constraints obtained using Planck. The tilt obtained by varying $C_\Theta$ falls within $n_\mathrm{s}\in[0.92,\,0.996]$, see Eq.~\eqref{Eq:Pzeta_ns_expressions}. Therefore the constraints on $C_\Theta$ can be prior dominated in some dataset combination where the tilt of the primordial power spectrum is not constrained well within this range. Lower prior on $m_\sigma/H$ is set to be the theoretical lower limit 1. In this case the clock field does not oscillate and we just get a smooth intermediate amplification in the slow-roll parameter resulting in suppression in the large scale power spectrum (the suppression is very similar to the suppression obtained in Whipped Inflation~\cite{Hazra:2014jka}). The higher limit is set to $m_\sigma/H=75$. Very high masses of the clock field will introduce spurious oscillations that are only supported for a very localized parameter space (usually extremely sharp oscillations are expected to fit noise in the data rather than addressing features and therefore remain highly localized) that is expected to be marginalized out. Also for very high masses $m_\sigma/H\sim100$, we notice instability in the numerical evolution. However the drop in the marginalized posterior distribution at high frequencies justifies our prior on the mass range. $\xi\sigma_f$ parameter is varied between 0 and 0.2 that allows sufficient amplitude of the sharp and resonant features for entire prior range of $m_\sigma/H$. The location of the feature parameter $N_0$ is allowed to vary within the range of the cosmological scales probed by Planck CMB ($\ell=2-2500$). We use linear priors on the $N_0$ parameter that effectively translates to logarithmic prior on the $k_0$ (the first mode for sharp feature signal) or $k_r$ (the first mode for resonant feature signal) used in~\cite{Chen:2014cwa}. Note that our prior volume is determined by theoretical limits and conservative observational windows that are expected to capture the global evidence of a model including the {\it look-elsewhere effect}.

$C_\sigma$ has subtle degeneracy with  $\xi\sigma_f$. Apart from marginal modification to the oscillation profile, $C_\sigma$ introduces an amplification of large scale power. For larger values of $C_\sigma\sim{\cal O}(0.1)$, a large $\xi\sigma_f$ suppresses the super-Hubble evolution of curvature perturbation and therefore we notice suppression at large scales; but for smaller $\xi\sigma_f$ values, we notice amplification. For very small values of $C_\sigma$ however, a very large $\xi\sigma_f$ introduces amplification at large scales. The interplay between the parameters dictates that for very small $C_\sigma$, $\sigma_f$ must be small to keep the same $m_\sigma/H$ which in turn demands a very large $\xi$ to keep the product $\xi\sigma_f$ same. A very large $\xi$ increases the coupling between curvature and isocurvature. The isocurvature sourcing increases the amplitude of curvature power at large scales. Since the effects of $C_\sigma$ is noticeable at large scales, we do not expect to constrain this parameter well and therefore we can expect one directional bound on this parameter where the amplification of the large scale angular power spectra is ruled out by the data. As can be noticed in the table, for $C_\sigma$ we allow a fairly conservative priors. 

Although in our main analyses we fix $\epsilon_i=10^{-7}$, we also investigate its variation in the nested sampling for completeness. In this case, also $\epsilon_i$ is varied together with the other parameters. Apart from controlling the tensor-to-scalar ratio $\epsilon_i$ amplifies the spectra at scales larger than the location of the oscillatory part of the feature, without impacting the latter, as explained in Ref.~\cite{Chen:2014cwa}. Naively this amplification should help constrain $\epsilon_i$ by placing an upper bound on it. However, its magnitude depends in a non-trivial way also on the combination of other parameters such as $C_\sigma$ and in particular $\xi\sigma_f$ which, in addition to governing the amplitude of the clock signal, also affects  isocurvature sourcing at large scales, as also discussed in Ref.~\cite{Braglia:2020fms}. A larger $\epsilon_i$ also enhances tensor modes, as it is accompanied by a larger inflationary energy scales. Nevertheless, for our prior choice ($\log_{10}\epsilon_i\in[-8,\,-4]$), their amplitude is well below the Planck+BK15 bound. We find the variation of $\epsilon_i$ increases the degeneracy within the model resulting in inefficient exploration of the prior volume. Since the parameter $N_0$ represents the position of features in the cosmological scales, we find that the posterior volume is concentrated on the low $N_0$ region, keeping the features at very large scales and thereby only preferring candidates that do not have features at small scales. Moreover opening up the parameter space to $\epsilon_i$ does not constrain the parameter and therefore fixing $\epsilon_i=10^{-7}$ does not limit our analysis. Any future detection of primordial B-modes will directly constrain this parameter that will necessitate a reanalysis of the model opening up $\epsilon_i$. 

\begin{table}
\centering
\begin{tabular}{|l|l|}
\hline
Parameters               & Priors      \\ \hline
$\Omega_\mathrm{b}h^2$   & $0.02,\, 0.0265$ \\ \hline
$\Omega_\mathrm{CDM}h^2$ & $0.1,\, 0.135$   \\ \hline
$100\theta$              & $1.03,\, 1.05$   \\ \hline
$\tau$                    & $0.3,\, 0.8$      \\ \hline
$C_\Theta$               & $0.002,\, 0.04$  \\ \hline
$\log_{10} P_{\zeta *}$              & $-8.7,\, -8$   \\ \hline
$C_{\sigma}$             & $0.005,\, 0.1$   \\ \hline
$m_\sigma/H$             & $1,\, 75$        \\ \hline
$\xi\sigma_f$            & $0,\, 0.2$       \\ \hline
$N_0$                    & $11,\, 16.5$     \\ \hline
\end{tabular}
\caption{\label{tab:priors} Priors on the background parameters and effective parameters used in our analysis. The priors on foreground and calibration parameters are kept unchanged from their default values in the Polychord analysis. In the analysis, we fix $\epsilon_i=10^{-7}$.}
\end{table}

\subsubsection{Finding best fit candidates}
Given the pool of the possible primordial spectra from the standard clock model, we concentrate on particular candidates that provide substantial improvement in fit compared to the baseline model. Depending on the frequency of oscillations, that is determined by the mass of the clock field $m_\sigma/H$, we characterize the candidates into low frequency (LFC -- $m_\sigma/H\in[1,\,25]$), medium frequency (MFC -- $m_\sigma/H\in[25,\,40]$) and high frequency (HFC -- $m_\sigma/H>40$) candidates. Given the large number of nuisance parameters, nested sampling or Markov-Chain Monte Carlo is not efficient to properly identify the improvement in fit to the data, though parameter estimation is not affected by this. We use BOBYQA~\cite{BOBYQA} algorithm to obtain the local and global best fit candidates in these 3 frequency domains. Given the multimodal parameter space for BOBYQA too, it is difficult to estimate the best fits. Therefore we use the samples from nested sampling and identify the {\it dips} in the $\chi^2$ {\it w.r.t.} the baseline best fit value. We identify the samples with better fits in the three frequency ranges. We then search for best fit using BOBYQA starting from the parameter values of the samples. Note that the total $\chi^2$ obtained from a particular data combination, for example TTTEEE+lowT+lowE can be broken down as follows:
\begin{equation}
    \chi^2_\mathrm{Total}=\chi^2_\mathrm{high-\ell~TTTEEE}+\chi^2_\mathrm{low-\ell ~TT}+\chi^2_\mathrm{low-\ell~EE}+\chi^2_\mathrm{prior} ~.
\end{equation}

Note that, here $\chi^2_\mathrm{prior}$ corresponds to the Gaussian priors on particular nuisance parameters. When a model is compared to the angular power spectra data, since at small scales foreground contributions become important, foreground angular spectrum is added to the `cosmological' lensed CMB power spectrum at each frequency channels. Then the cosmological and nuisance parameters are varied together for the model comparison against the data. In addition to the $\chi^2$s from individual data, $\chi^2_\mathrm{prior}$s are added. As discussed in Table 16 of the Planck Likelihood analysis paper~\cite{Aghanim:2019ame}, the Plik likelihood uses Gaussian priors on power spectrum calibration for the 100GHz, 200GHz and on the absolute map calibration parameters. At the same time we also use default Gaussian priors on the amplitudes of Galactic dust power at different maps (TT, TE, EE) and at different frequencies. Moreover, using the information on kinetic and thermal Sunyaev-Zeldovich power spectrum (${\cal D}^{SZ}$) obtained from the CMB observations by Atacama Cosmology Telescope~\footnote{\href{https://act.princeton.edu/}{https://act.princeton.edu/}} and South Pole Telescope~\footnote{\href{https://pole.uchicago.edu/spt/}{https://pole.uchicago.edu/spt/}}, Planck likelihood imposes Gaussian priors on the combination of SZ power spectrum. 

Search for best fit optimizes the $\chi^2_\mathrm{Total}$ by the variation of the nuisance parameters. We explore three methods in order to explore the best fits that minimize the $\chi^2_\mathrm{Total}$ and only CMB $\chi^2$. 

\paragraph{Method-I: Fixed nuisance parameters} In this method we explore the best fit by fixing the nuisance parameters to the best fit nuisance parameters obtained from the baseline power law $\Lambda$CDM best fit to the data. We vary only 4 background and 6 inflation parameters starting from the {\it dips} in the samples in $\chi^2$ plane. In this case the $\chi^2_\mathrm{prior}$ remains the same as in the baseline model.  

\paragraph{Method-II: Neglecting $\chi^2$ for nuisance priors} Here we impose uniform priors on all the nuisance parameters and therefore $\chi^2_\mathrm{prior}$ does not contribute to the $\chi^2_\mathrm{Total}$. However, unlike method-I, here we vary all the nuisance parameters alongside the cosmological parameters. Here we just optimize CMB $\chi^2$ allowing the freedom in both CMB and foreground power spectrum. Note that here $\Delta\chi^2=\chi^2_\mathrm{model}-\chi^2_\mathrm{baseline}$ uses the $\chi^2_\mathrm{baseline}$ also obtained in method-II. In this method the $\chi^2_\mathrm{Total}$ for any model will be the lowest as we have maximum freedom in the variation of parameters without prior constraints. 

\paragraph{Method-III: Optimizing $\chi^2_\mathrm{Total}$} In this method, we optimize the $\chi^2_\mathrm{Total}$ by the variation of cosmological background, inflation and nuisance parameters including the Gaussian priors on the nuisance parameters. The calculation of $\Delta\chi^2$ uses same $\chi^2_\mathrm{baseline}$ as in method-I but $\chi^2_\mathrm{model}$ exploits the freedom of variation of the nuisance parameters. However, since the prior constraint in the nuisance parameter is included, this method is the most conservative one and should be used for reporting improvement in fit to the data. $\Delta\chi^2$  in this method should be between the ones obtained in method-I and method-II. 

We find the best fit cosmological (background+inflationary) parameters obtained between these methods differ marginally which assures that the variation in the foreground parameters does not alter the best fit candidates significantly. In this paper, for simplicity, we will present all the residual plots in method-III. While for P18 we find the best fits in all three methods, for EG20 we only use method-III because of fewer nuisance parameters.

\begin{figure}
\includegraphics[width=\columnwidth]{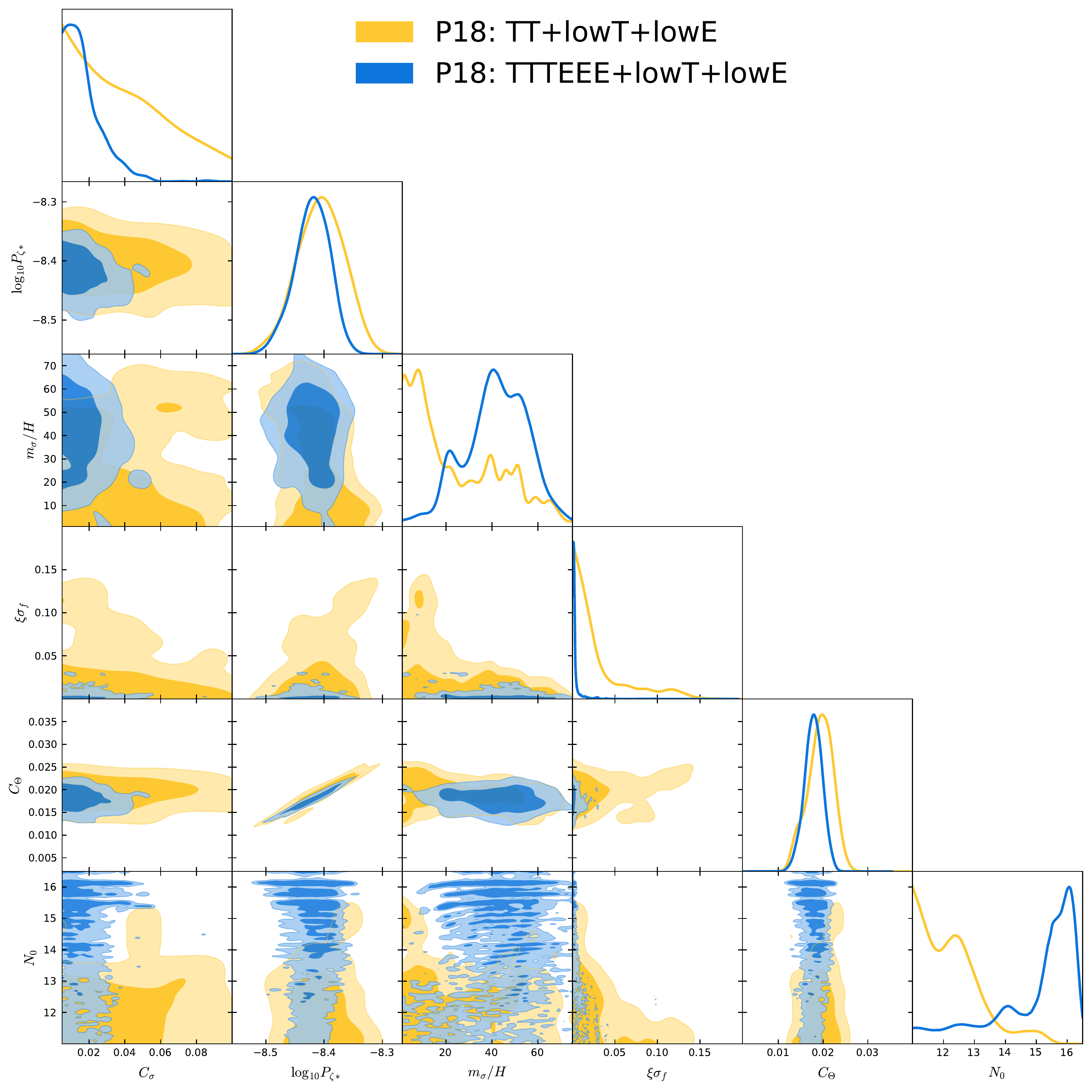}
\caption{\footnotesize\label{PSC-6P-Bin1-TTTEEE} Triangle plot for PlikHM bin1 (P18) TT+lowT+lowE and TTTEEE+lowT+lowE datasets. Only constraints on the CPSC model parameters are plotted.}
\end{figure}

\begin{figure}
\includegraphics[width=\columnwidth]{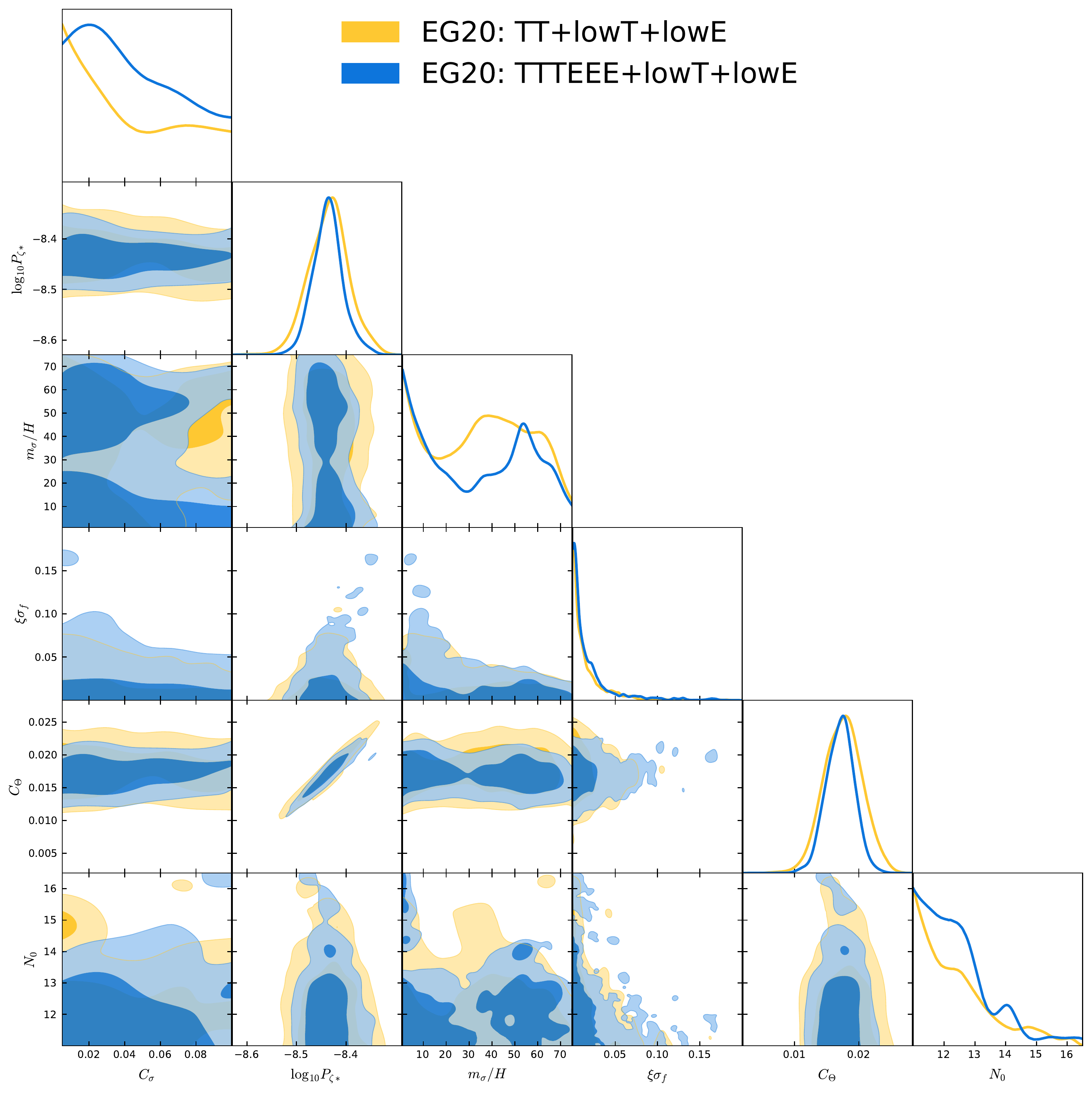}
\caption{\footnotesize\label{PSC-6P-EG20-TTTEEE} Triangle plot for Clean CamSpec v12.5HMcln (EG20) TT+lowT+lowE and TTTEEE+lowT+lowE datasets. Only constraints on the CPSC model parameters are plotted.}
\end{figure}

\begin{figure}[t!]
\includegraphics[width=\columnwidth]{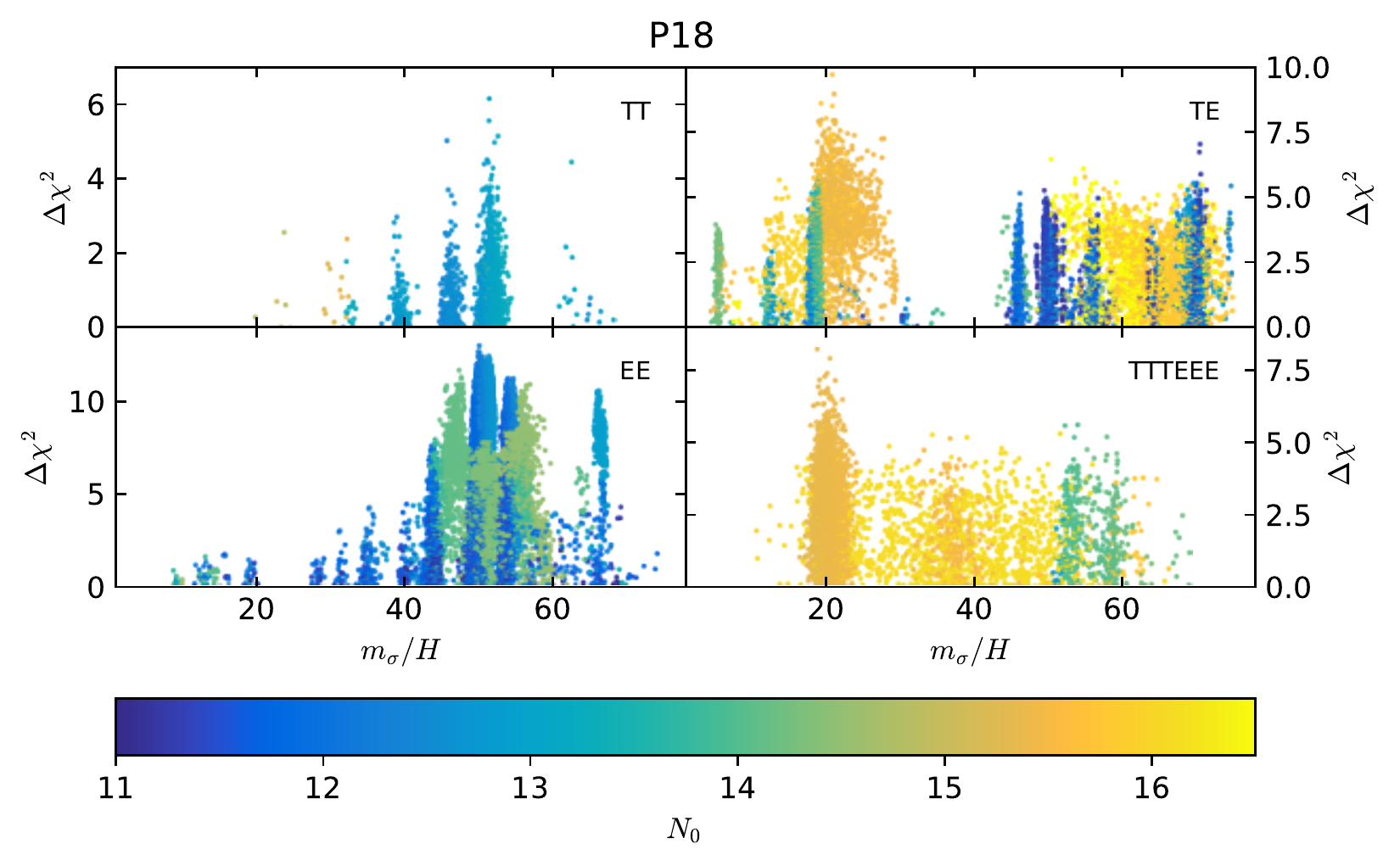}
\caption{\footnotesize\label{PSC-6P-P18-deltachisquare} Improvement in the fit to P18 compared to the baseline is plotted for samples collected from nested sampling. $\Delta\chi^2$ is plotted against the parameter $m_\sigma/H$ and $N_0$.}
\end{figure}

\begin{figure}[b!]
\includegraphics[width=\columnwidth]{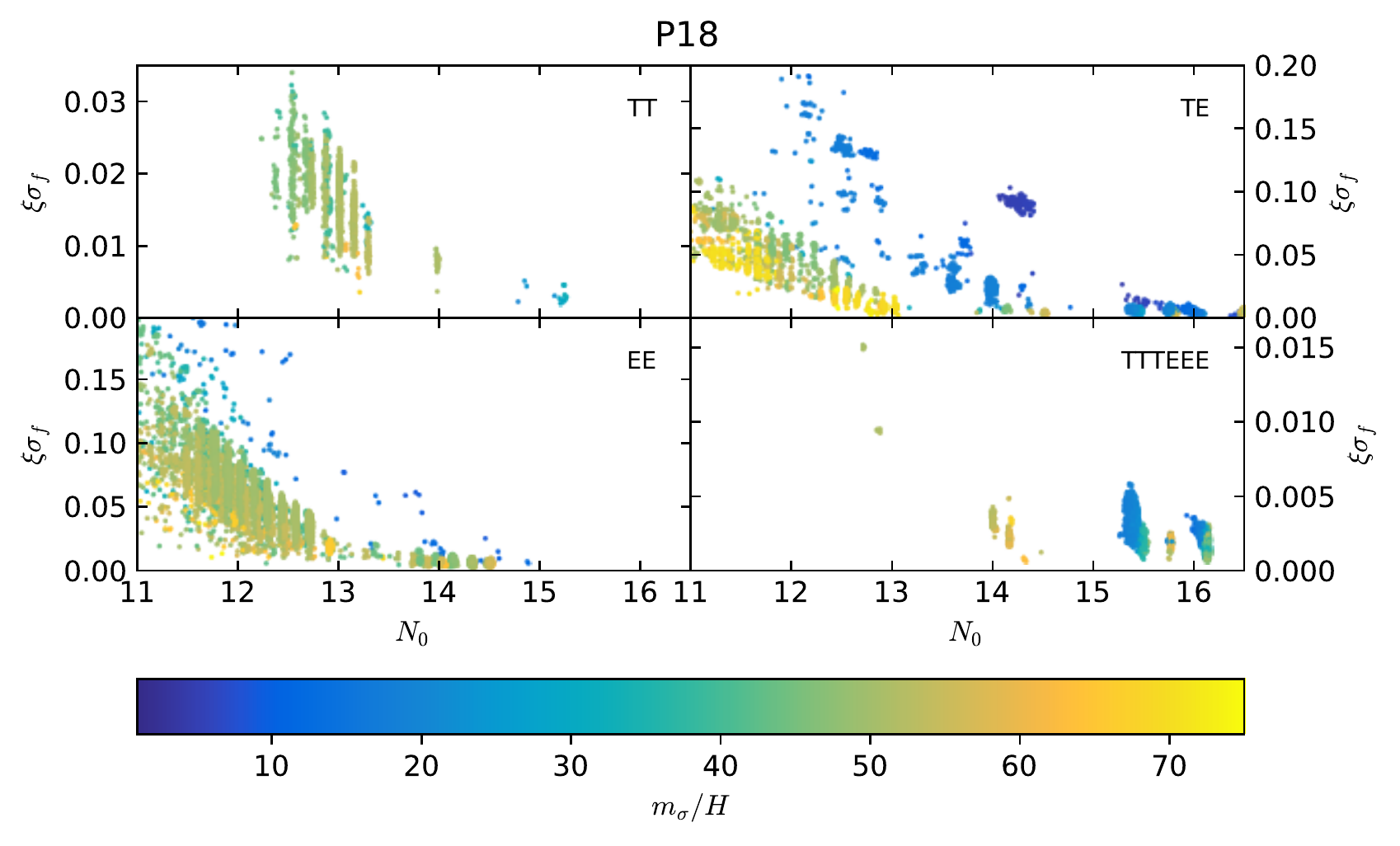}
\caption{\footnotesize\label{PSC-6P-P18-heatmap} All the samples obtained (using P18) that are better than the baseline model are plotted here. Similar positions and colors in different plots indicate same candidates providing improvement in fit to different datasets.}
\end{figure}

\clearpage

\begin{figure}[t!]
\includegraphics[width=\columnwidth]{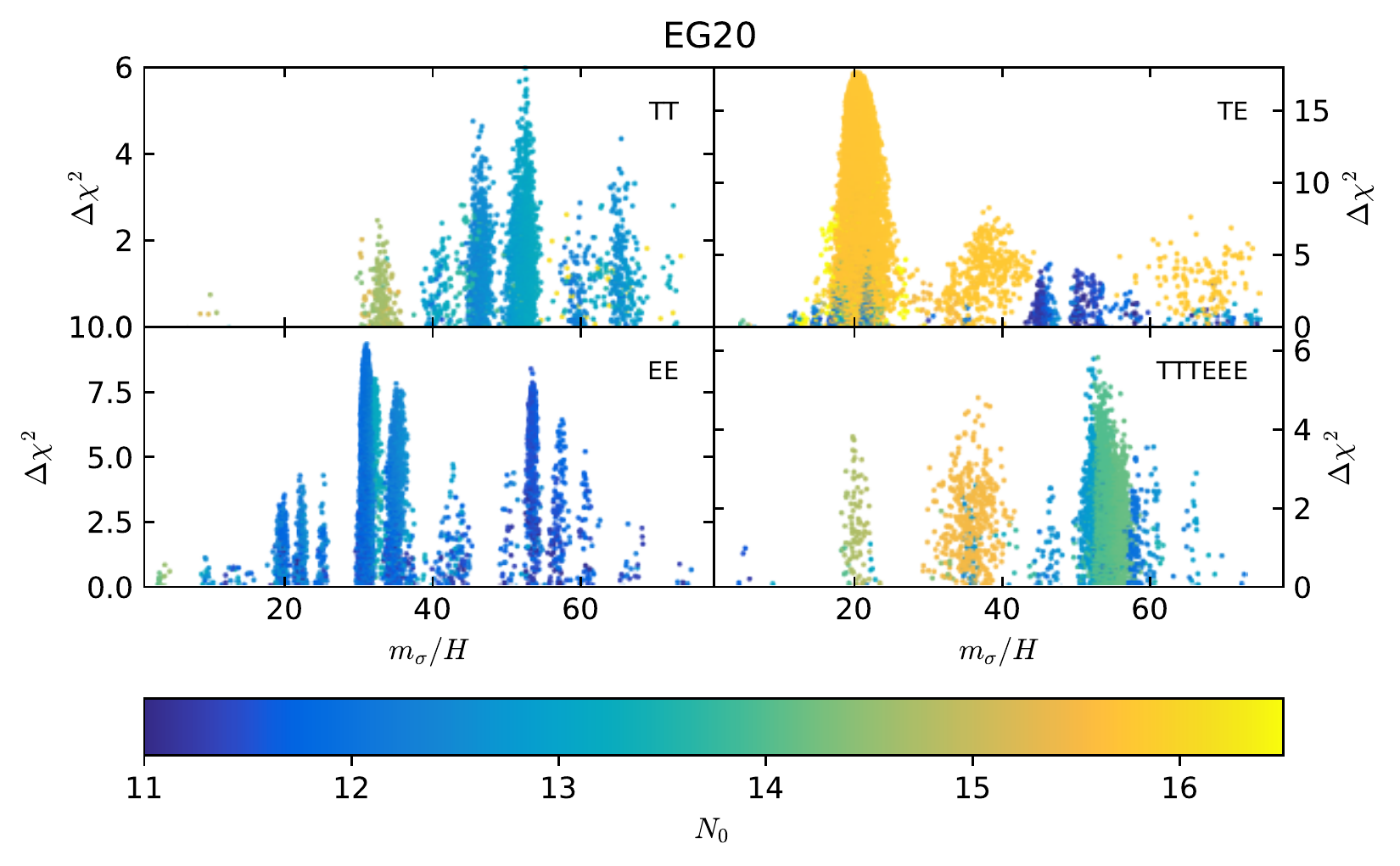}
\caption{\footnotesize\label{PSC-6P-EG20-deltachisquare} Same as in Fig.~\ref{PSC-6P-P18-deltachisquare} but for Clean CamSpec v12.5HMcln (EG20) likelihood.}
\end{figure}

\begin{figure}[b!]
\includegraphics[width=\columnwidth]{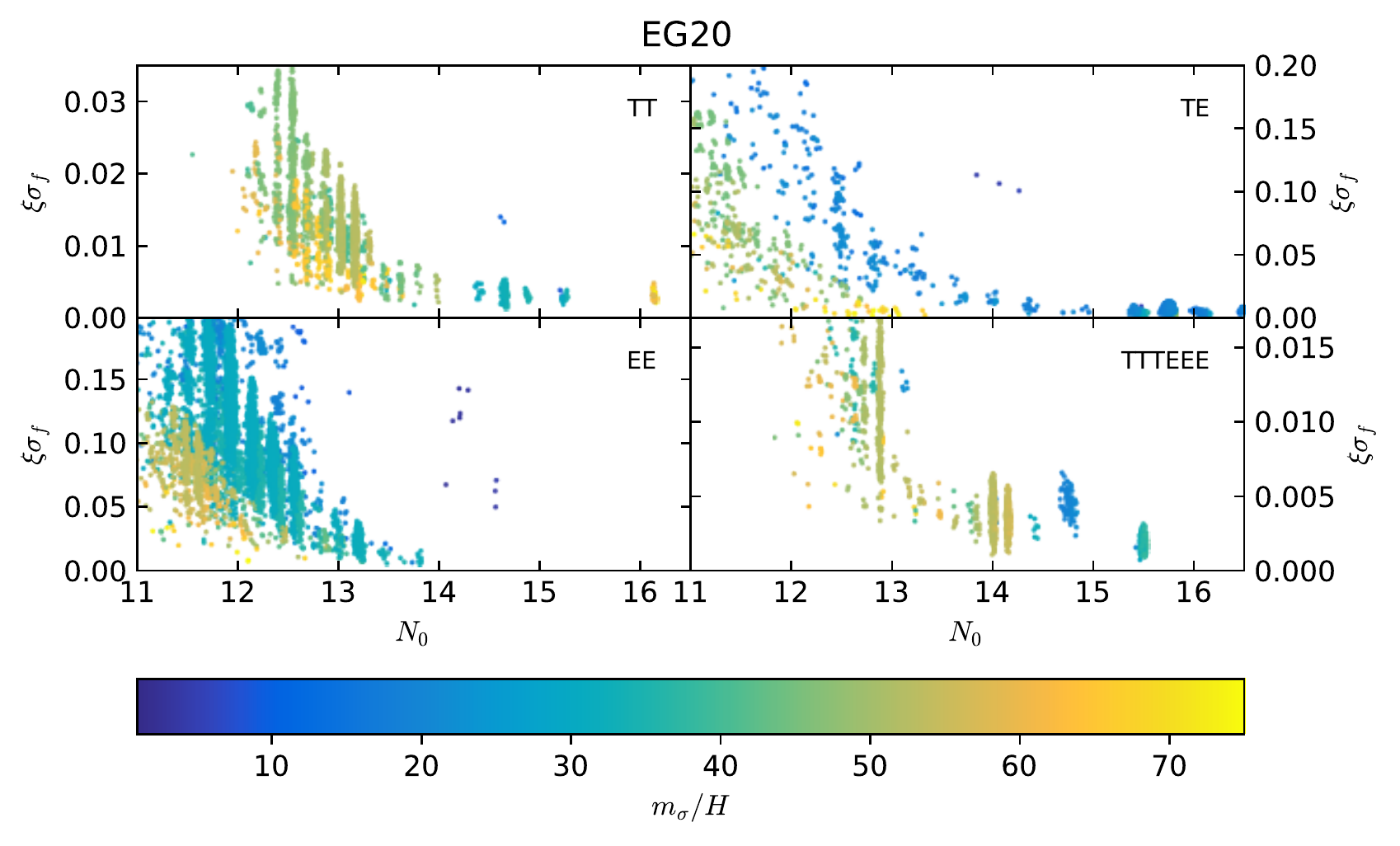}
\caption{\footnotesize\label{PSC-6P-EG20-heatmap} Same as in Fig.~\ref{PSC-6P-P18-heatmap} but for Clean CamSpec v12.5HMcln (EG20) likelihood.}
\end{figure}

\clearpage

\begin{figure}[t!]
\includegraphics[width=\columnwidth]{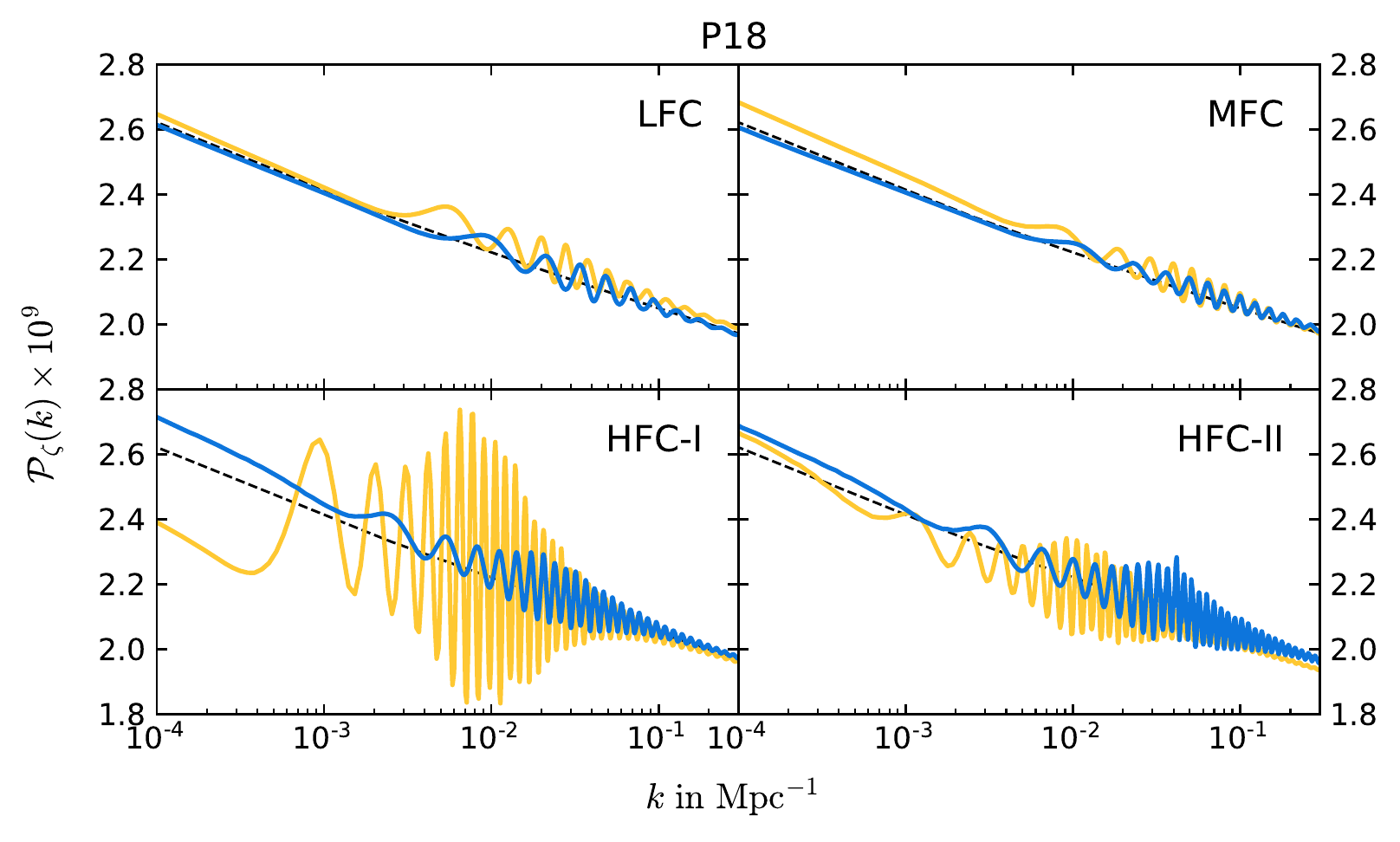}
\caption{\footnotesize\label{fig:PPS-P18-All} Best fit primordial power spectra obtained from P18. We plot the best spectra that are tabulated in Tab.~\ref{tab:candidates}. Top panel shows the LFC (left) and MFC (right) while the bottom panel contains the HFC-I (left) and HFC-II (right). We use the same choice of colors of Fig.~\ref{PSC-6P-Bin1-TTTEEE}, i.e. gold for TT+lowT+lowE and blue for TTTEEE+lowT+lowE.}
\end{figure}

\begin{figure}[b!]
\includegraphics[width=\columnwidth]{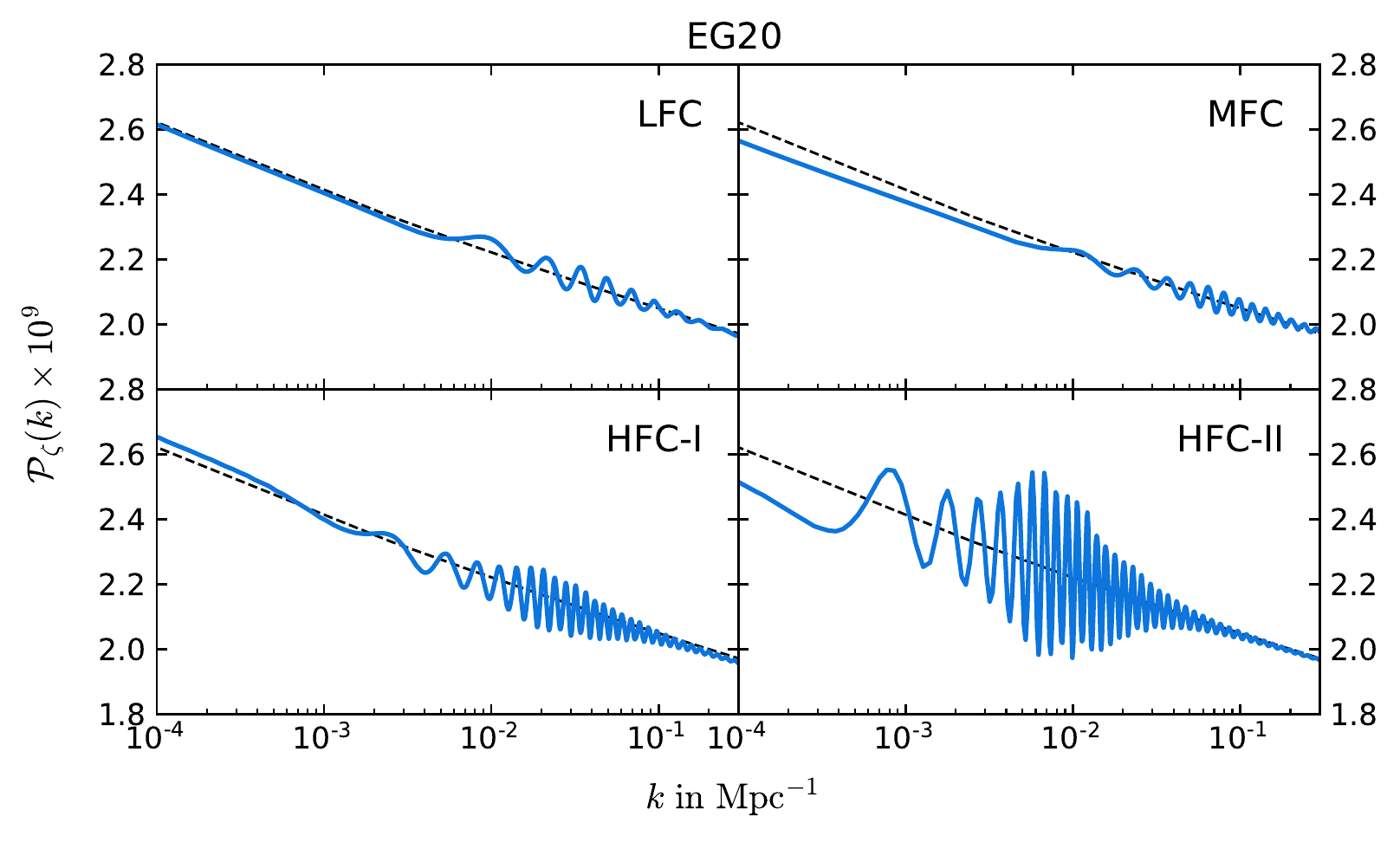}
\caption{\footnotesize\label{fig:PPS-EG20-All} Best fit primordial power spectra obtained from EG20. We plot the best spectra that are tabulated in Tab.~\ref{tab:candidates}. Top panel shows the LFC (left) and MFC (right) while the bottom panel contains the HFC-I (left) and HFC-II (right).}
\end{figure}

\clearpage

\section{Results}~\label{Sec:ComparePlanck}
\setcounter{equation}{0}

We now present the results of our data analysis. Although our ultimate goal is to find bestfit candidates that improve the fit to data {\it w.r.t.} to the baseline model, it is very important to discuss the statistics of our nested sampling. This will also help us understand the regions of the parameter space of the CPSC
model that better fit the Planck 18 data.

\subsection{Parameter estimation and marginal likelihood}
We show the triangle plots of the posterior distribution of the CPSC parameters  in Figs.~\ref{PSC-6P-Bin1-TTTEEE} and \ref{PSC-6P-EG20-TTTEEE} for P18 and EG20 respectively\footnote{The plots are obtained using the GetDist package \cite{Lewis:2019xzd}.}.  In order not to overcrowd our plots, we only plot two of the data combinations considered, i.e. TT+lowT+lowE and TTTEEE+lowT+lowE. As expected, the only parameters that are well constrained and show a posterior very close to be Gaussian are $\log_{10} P_{\zeta *}$ and $C_\Theta$, which govern the amplitude of the PPS  and its tilt (usually parameterized by $A_s$ and $n_s$ respectively), and are very tightly constrained in the baseline model. Here, unlike $A_s$ and $n_s$ (compare with Fig.~2 of the Planck Inflation paper \cite{Akrami:2018odb}), the parameters $\log_{10} P_{\zeta *}$ and $C_\Theta$ show a small degeneracy due to the fact that $C_\Theta$ slightly affects the amplitude too, as discussed in Section \ref{Sec:Method}.

The remaining 4 parameters are not constrained as well as the former ones and their posterior distributions are slightly different for P18 and EG20. However, it is interesting to see that both  set an upper bound on $\xi\sigma_f$, which represents the amplitude of the clock oscillations, and on $C_\sigma$. Indeed, as discussed in the previous section, a large $C_\sigma$ can generate amplification of power at large scales, which is not supported by data. 
The most interesting posterior distribution, however, is the one for $m_\sigma/H$, which encodes valuable information on the frequencies of the clock signals.   As can be seen from the triangle plots, several frequencies ranging from LFC to HFC are supported by data. We stress that this is the first work where the frequency of such signals is allowed to vary in the data analysis.

Because of the multimodal nature of the posteriors, in particular those of $m_\sigma/H$ and $N_0$, however, it is not easy to appreciate all the information provided by the nested sampling by simply looking at the triangle plots. To better visualize our results, we show the scatter plots of the samples that provide a better fit than the baseline model. Specifically in Fig.~\ref{PSC-6P-P18-deltachisquare}, for P18, we plot the $\Delta\chi^2$ distribution as a function $m_\sigma/H$ using different colors depending on the parameter $N_0$, governing the location of the feature,  and in Fig.~\ref{PSC-6P-P18-heatmap} the distribution of samples in the  $\xi\sigma_f$ vs $N_0$ plane  with $m_\sigma/H$ used as colorbar. The same plots for EG20 are presented in Figs.~\ref{PSC-6P-EG20-deltachisquare} and \ref{PSC-6P-EG20-heatmap}. These plots are also useful to understand the fit of our model to each of the 4 combinations of data considered.

First of all, we notice that for all 4 combinations, except for TE+lowE, the average $\Delta\chi^2$ is smaller for EG20 compared to P18, confirming the constraining power claimed in \cite{Efstathiou:2019mdh}.

Moving on to comment on each dataset, for TT+lowT+lowE in both P18 and EG20, we clearly see three dense regions that improve the fit to data around $m_\sigma/H\sim 40,\,45,\,52$ and also less dense regions around $m_\sigma/H\sim 33,\,60,\,66$. The improvement in the fit occurs for features at relatively large scales, i.e. $N_0\in[12,\,14]$, corresponding to the sharp feature part of the full signal starting at scales $k_0\in[0.6,\,4.6]\times 10^{-4}$ Mpc$^{-1}$, except for the region around $m_\sigma/H\sim 33$ for which $N_0\sim 15$ which corresponds to $k_0 \sim 10^{-3}$ Mpc$^{-1}$.

Concerning EE+lowE, we notice on average a larger improvement in $\Delta\chi^2$ compared to the previous combination. We see that now P18 data present different stripes in the $\Delta\chi^2$ vs $m_\sigma/H$ plane, whereas for EG20 samples predominantly gather in the HFC region.
In both cases, features with $N_0\lesssim14$ are preferred, which all have low $k_0$ values. Because with the current precision of Planck, TT+lowT+lowE has more constraining power than EE+lowE for the type of oscillatory signals studied here, these large amplitude candidates would be strongly disfavored once jointly analyzed with TT+lowT+lowE.  Therefore, in our study, we will not present any candidates that are picked up by EE+lowE alone.

In the case of TE+lowE, several frequencies can improve the fit {\it w.r.t.} the baseline model. However, we notice many samples that pile up in the LFC region around $m_\sigma/H\sim 20$ for large values of $N_0$ in both P18 and EG20. The heatmaps in Figs.~\ref{PSC-6P-P18-heatmap} and \ref{PSC-6P-EG20-heatmap} suggest that these points have a very small $\xi\sigma_f$ (small amplitude of oscillations) which together with a very large $N_0$ (very small scales feature), indicate a fit similar to a power-law PPS.

The results with the full TTTEEE+lowT+lowE likelihoods are significantly more stringent and select localized regions in the parameter space of the CPSC model, shown in the bottom-right corners of Figs.~\ref{PSC-6P-P18-deltachisquare}, \ref{PSC-6P-EG20-deltachisquare}, \ref{PSC-6P-P18-heatmap} and \ref{PSC-6P-EG20-heatmap}.  These regions will be the starting point of the bestfit analysis of the next Subsections.

Another important point of our nested sampling procedure is that it allows us to compute the Bayesian evidence for our model. As discussed, this is crucial for model comparison and is a measure of the extent to which the improvement in the fit to data (extensively discussed below) is enough to warrant the addition of the 4 extra parameters of this particular model. For each datset, we report the Bayes factor, defined as $\ln B\equiv\ln {\cal B}_{\rm CPSC}-\ln {\cal B}_{\rm baseline}$, in Table~\ref{tab:evidence}. Here the baseline model refers to the featureless subset of CPSC models where we keep the 4 feature parameters $C_\sigma,\,m_\sigma/H,\,\xi\sigma_f$ and $N_0$ fixed such that we get power law spectrum where only the amplitude and tilt are varied with $\log_{10} P_{\zeta *}$ and $C_\Theta$. Apart from fixing feature parameters, we keep the same prior for other parameters in the baseline model as in the CPSC model that ensures the additional prior volume in CPSC model only comes from the features.
Since this factor integrates over all prior ranges, it also addresses the look-elsewhere effect that is hard to address by the individual $\Delta\chi^2$ value of each best-fit candidate.
With this definition, a negative Bayes factor, as is our case, corresponds to a preference of the baseline model over the  CPSC one. According to the Jeffreys' scale, the baseline model is \emph{weakly} preferred over this CPSC model.

\renewcommand{\arraystretch}{1.5}
\begin{table}
\centering
\begin{tabular}{|l|l|l|}
\hline
Dataset          & P18 (Plik bin1) & EG20 (CamSpec v12.5HMcln) \\ \hline
TT+lowT+lowE     &      -1.76           & -2.28                     \\ \hline
TE+lowE          &      -1.35           & -0.93                     \\ \hline
EE+lowE          &      -1.28           &-1.00                        \\ \hline
TTTEEE+lowT+lowE &      -1.30           & -2.00                     \\ \hline
\end{tabular}
\caption{Bayes factors ($\ln B$) obtained for the CPSC model {\it w.r.t.} the baseline model. The baseline model has the same prior volume as in the featureless part of the CPSC model. The second column in the table presents the Bayes factors for the datasets combinations where high-$\ell$ data is used from PlikHM unbinned bin1 data (P18) and the third column corresponds to the Clean CamSpec (EG20) high-$\ell$ temperature and polarization data. }
\label{tab:evidence}
\end{table}

\begin{table}
\begin{center}
\begin{tabular}{|c||c|c|c|c||c|c|c|c|}
\hline
                     & \multicolumn{4}{c||}{P18 (Plik bin1)} & \multicolumn{4}{c|}{EG20 (CamSpec v12.5HMcln)} \\ \hline
                     & LFC     & MFC     & HFC-I   & HFC-II & LFC        & MFC       & HFC-I     & HFC-II    \\ \hline
$\xi$           & 0.276  & 0.285  & 0.429   & 0.530 & 0.269     &  0.294   & 0.361    & 1.084    \\ \hline
$\sigma_f(\times10^{-2})$           & 1.370  & 0.700  & 1.029   & 0.802 & 1.288     & 0.658    & 1.109    & 1.086    \\ \hline
$V_\mathrm{inf}\,(\times10^{-14})$           & 9.162  & 8.898  & 9.107   & 8.899 & 9.191     & 8.595    & 8.548    & 8.458    \\ \hline

$C_\sigma$           & 0.013   & 0.0090   & 0.050    & 0.050   & 0.011       & 0.0097    & 0.058     & 0.054     \\ \hline
$C_\Theta$           & 0.0187  & 0.0175  & 0.0190   & 0.0185 & 0.0187     & 0.0166    & 0.0174    & 0.0175    \\ \hline \hline
$N_{0}$   & 15.39   & 15.48   & 14.00      & 14.19  & 15.40       & 15.50      & 14.00        & 12.89     \\ \hline 
$\log_{10} P_{\zeta *}$        & -8.409  & -8.423  & -8.411  & -8.421 & -8.408     & -8.439    & -8.440     & -8.437    \\ \hline
${m_\sigma}/{H}$ & 20.24   & 33.28   & 53.39   & 68.63  & 20.26      & 36.78     & 52.97     & 52.35     \\ \hline
$\xi\sigma_f$        & 0.0038  & 0.0020   & 0.0044  & 0.0043 & 0.0035     & 0.0019    & 0.0040     & 0.0120     \\ \hline\hline
$\Delta\chi^2$       & 11.4        &     6.8   &     8.8   & 15.7      &     9.7    &    7.5       &    8.3       &     8.5      \\ 
       & (13.5)        &     (9.4)    &     (11.7)    & (19.1)       &         &           &           &           \\ \hline
$k_0\,(\times10^{-3})$       &  1.857       &    2.024     &  0.468       & 0.569       &   1.879         &   2.061        &    0.468 &     0.155      \\ 
$\mathrm{Mpc}^{-1}$       &        &     &        &       &            &           &     &        \\ \hline

$k_r\,(\times10^{-2})$       &  3.740       &      6.735   &     2.497    & 3.907       &     3.805       &    7.579       &     2.479      &    0.81       \\ 
$\mathrm{Mpc}^{-1}$       &        &     &        &       &            &           &     &        \\ \hline

\end{tabular}
\caption{Best fit candidates found in P18 and EG20 likelihoods. Depending on the frequency of the clock signal ($m_\sigma/H$) the candidates fall in three categories: low, medium and high-frequency regime. In the high frequency regime we have two different best fits (HFC-I and II). In P18 likelihood HFC-I and II correspond to two different high frequency features. However, for EG20 likelihood they correspond to features with similar frequencies (similar ${m_\sigma}/{H}$) but located at different scales (different $N_0$). The first 5 rows in the table contain the model parameters that appear in the Lagrangian in Eq.~\eqref{eq:LagrNew}. 
In the next 4 rows we present the parameters that are derived from the parameters appearing in the Lagrangian. 
The difference in $\chi^2$ are quoted {\it w.r.t.} the power law $\Lambda$CDM model with method III. In parenthesis, for P18 likelihoods we provide the improvement in method II as well. As mentioned in the text, improvement in method III should be within the improvements obtained in method-I and method-II.}
\label{tab:candidates}
\end{center}
\end{table}

\subsection{Best fit candidates}

In Table~\ref{tab:candidates} we list the best-fit parameter values and improvements {\it w.r.t.} the power law model for both P18 and EG20 likelihoods. We provide the model parameters as they appear in the Lagrangian Eq.~\ref{eq:LagrNew}, followed by the effective parameters (see discussions in Sec.~\ref{subsec:effective_parameters}). Following the improvement in fit compared to the baseline model, we tabulate the modes $k_0$ and $k_r$, the first mode in the sharp feature and resonant feature signal, respectively. These two modes are related as $k_r=(m_\sigma/H) k_0$ \cite{Chen:2011zf,Chen:2014cwa}. For both likelihoods we divide the best fit candidates in three frequency regions: low, medium and high frequencies. In high frequency region we present two feature candidates that are qualitatively different in their frequency and/or scales. We mainly discuss candidates for TTTEEE+lowE+lowT, that we plot in Figs.~\ref{fig:PPS-P18-All} and in~\ref{fig:PPS-EG20-All} in blue for P18 and EG20 respectively. For P18,  we also compare these candidates  with those with similar frequencies from TT+lowT+lowE, that we plot in Fig.~\ref{fig:PPS-P18-All} in gold.

\subsubsection{Low frequency candidates}

\paragraph{TTTEEE candidates:}

\begin{figure}
\centering
\includegraphics[width=0.96\columnwidth]{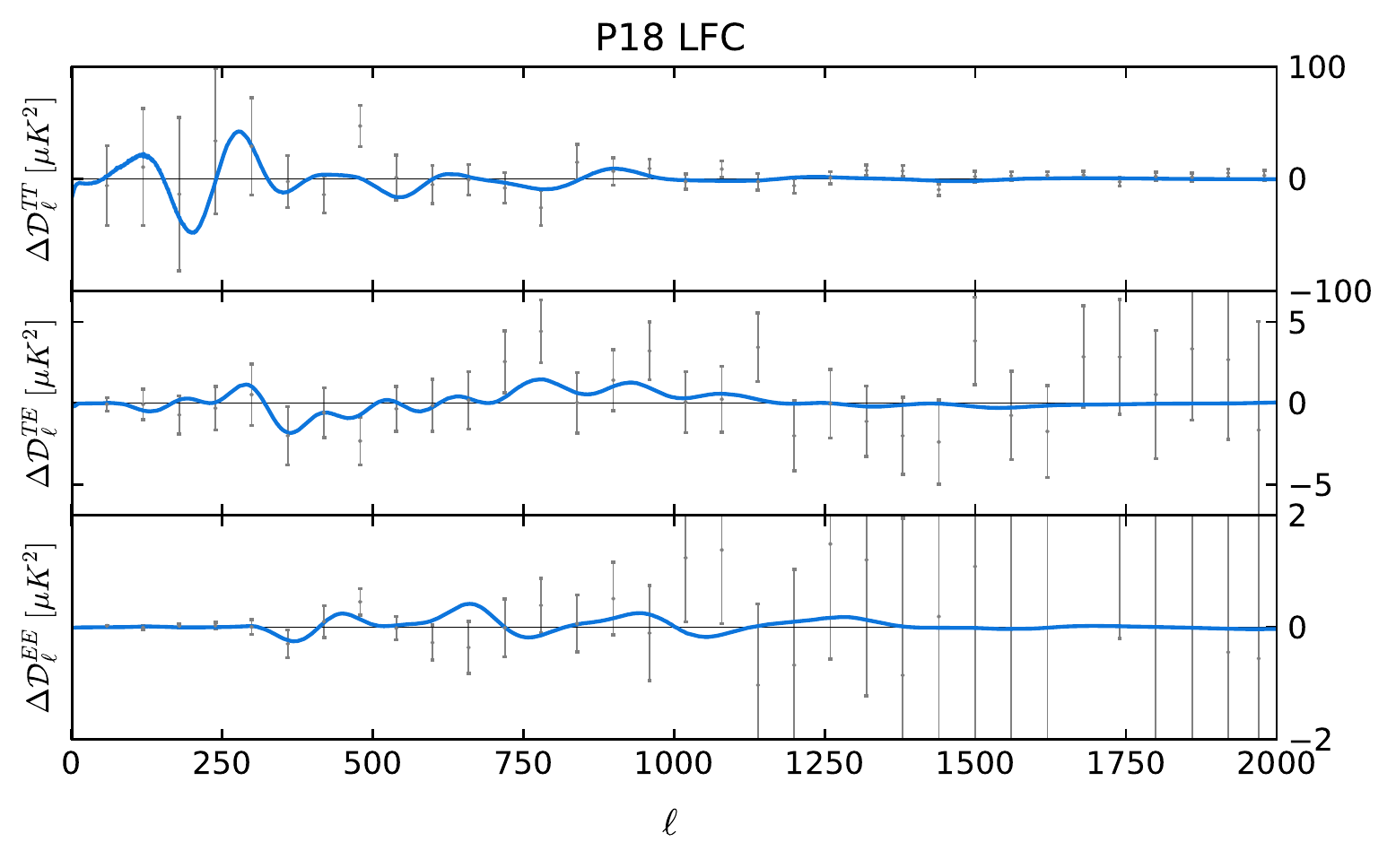}
\caption{\footnotesize\label{fig:Resid-18-LFC} Residual plots in the standard clock model for low frequency TTTEEE best fit to Plik bin1 likelihood. We find $\Delta\chi^2=11.4$ compared to baseline in method III.}
\end{figure}

\begin{figure}
\centering
\includegraphics[width=0.96\columnwidth]{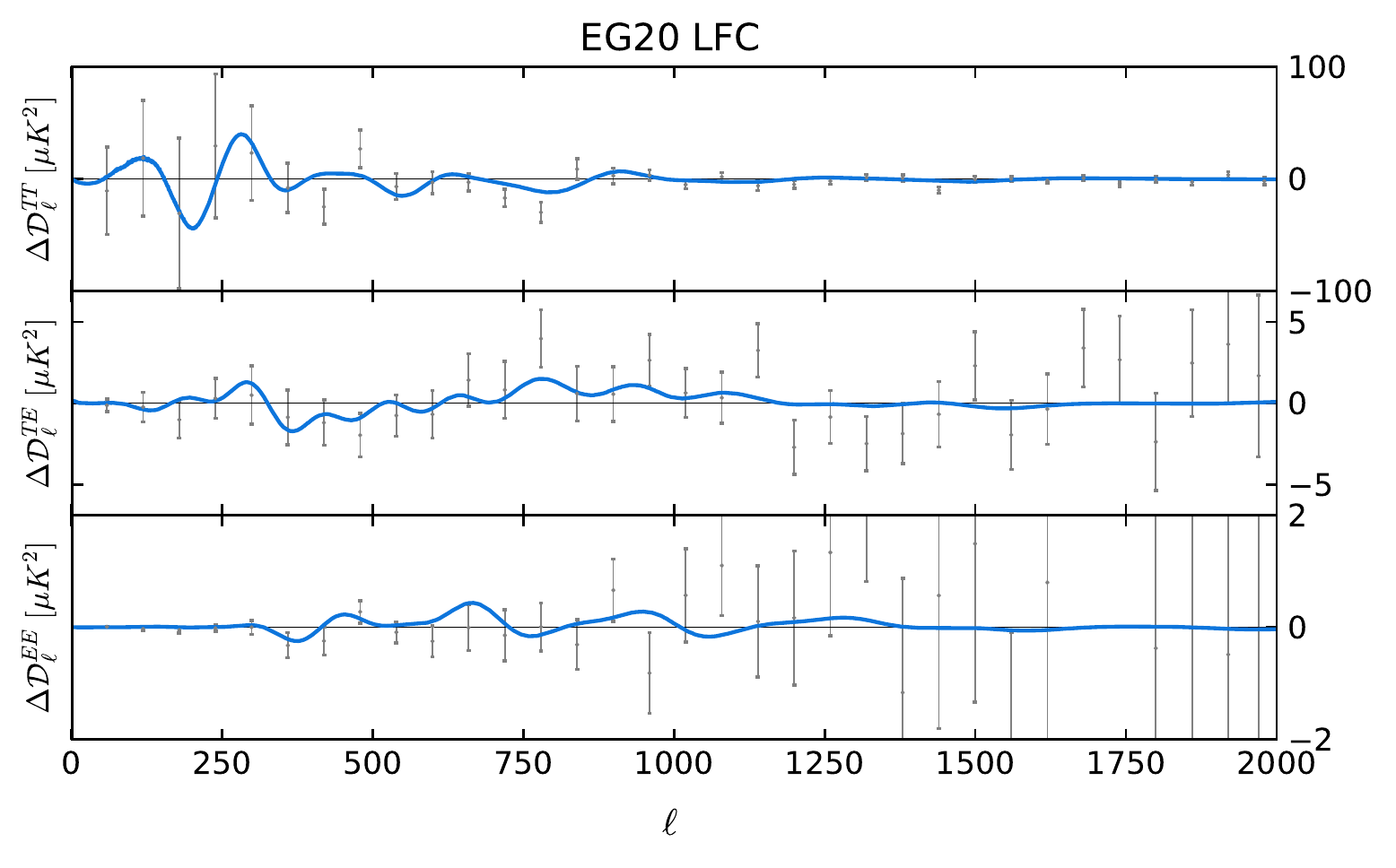}
\caption{\footnotesize\label{fig:Resid-EG20-LFC} Residual plots in the standard clock model for low frequency TTTEEE best fit to EG20 likelihood. We find $\Delta\chi^2=9.7$ compared to baseline in method III.}
\end{figure}

Best fit LFC for both P18 and EG20 are obtained around $m_\sigma/H\sim20$. We obtain 11 improvement in fit to the $\chi^2$ in P18 compared to baseline in method-III. In method-II, we get even more improvement, but as the nuisance prior likelihood is not included there, the result is not conservative. In EG20 the improvement goes down marginally and we get an overall 10 improvement in $\chi^2$ there. At large scales we do not get any suppression in the LFC. The angular power spectra that are residual to the baseline best fits are plotted in Figs.~\ref{fig:Resid-18-LFC} and~\ref{fig:Resid-EG20-LFC} for P18 and EG20 respectively. We also plot the binned data, residual to the baseline best fit. Here we use a binwidth of $\Delta\ell=60$. The plots indicate that LFC provide wide features that fit outliers that are broad in nature. The features address the residual oscillations around the first acoustic peak and the dip followed by a bump around $\ell\sim750-850$ in the TT data (Note that  reconstruction of primordial spectrum revealed this $\ell\sim750-850$ feature~\cite{Hazra:2014jwa} in Planck 2013 TT data). The same primordial spectrum can be noticed to fit the dip at $\ell\sim300-500$ and the bump around $\ell\sim750$ in the TE residual. Even though EE has larger uncertainties, the LFC seem to address the dip and bump at $\ell\sim300-500$ in the EE residual. We also find that the parameter space around the LFC is favored by both TT and TE data in the sample plots in Figs.~\ref{PSC-6P-P18-deltachisquare} and~\ref{PSC-6P-P18-heatmap}. Generally speaking, same spectrum having support from different datasets increases the possibility of this feature being a {\it true feature} rather than statistical fluctuations. Increase in the sky fraction in EG20 only marginally reduces the fit to the data by the feature.  

\paragraph{A comparison with TT candidates:}
The P18 TT+lowT+lowE best fit candidate around the same $m_\sigma/H$ range as LFC is plotted in gold in Fig.~\ref{fig:PPS-P18-All}. This best fit candidate has features located at larger scales compared to the LFC with increased amplitude while the oscillations at the small scales have decent overlap. When we compare with the full data the overall improvement in $\chi^2$ obtained from the full covariance matrix is minimized and therefore the best fit relocates from TT best fit candidate.

\begin{figure}
\centering
\includegraphics[width=0.96\columnwidth]{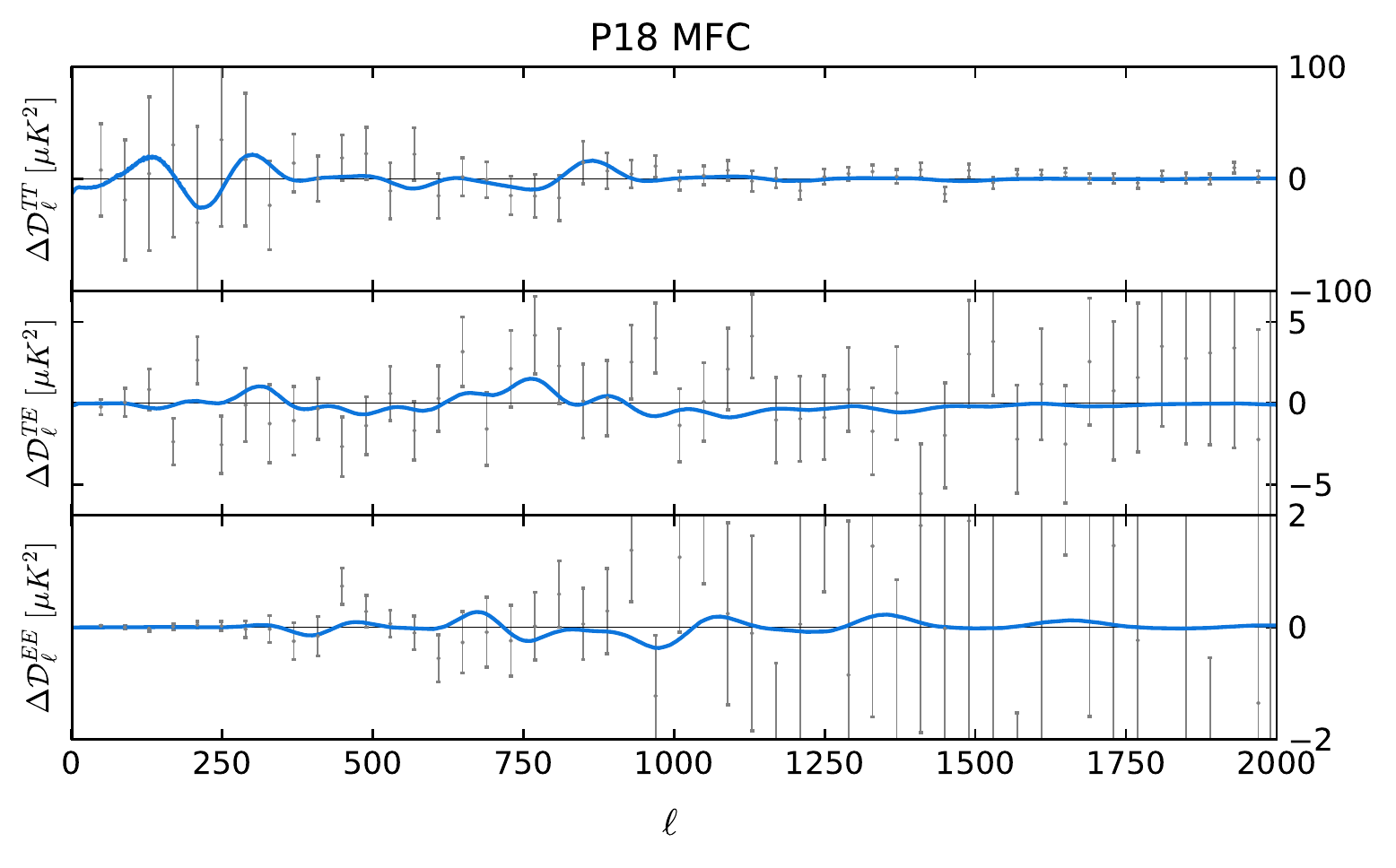}
\caption{\footnotesize\label{fig:Resid-P18-MFC} Residual plots in the standard clock model for medium frequency TTTEEE best fit to P18 likelihood. We find $\Delta\chi^2=6.8$ compared to baseline in method III.}
\end{figure}

\begin{figure}
\centering
\includegraphics[width=0.96\columnwidth]{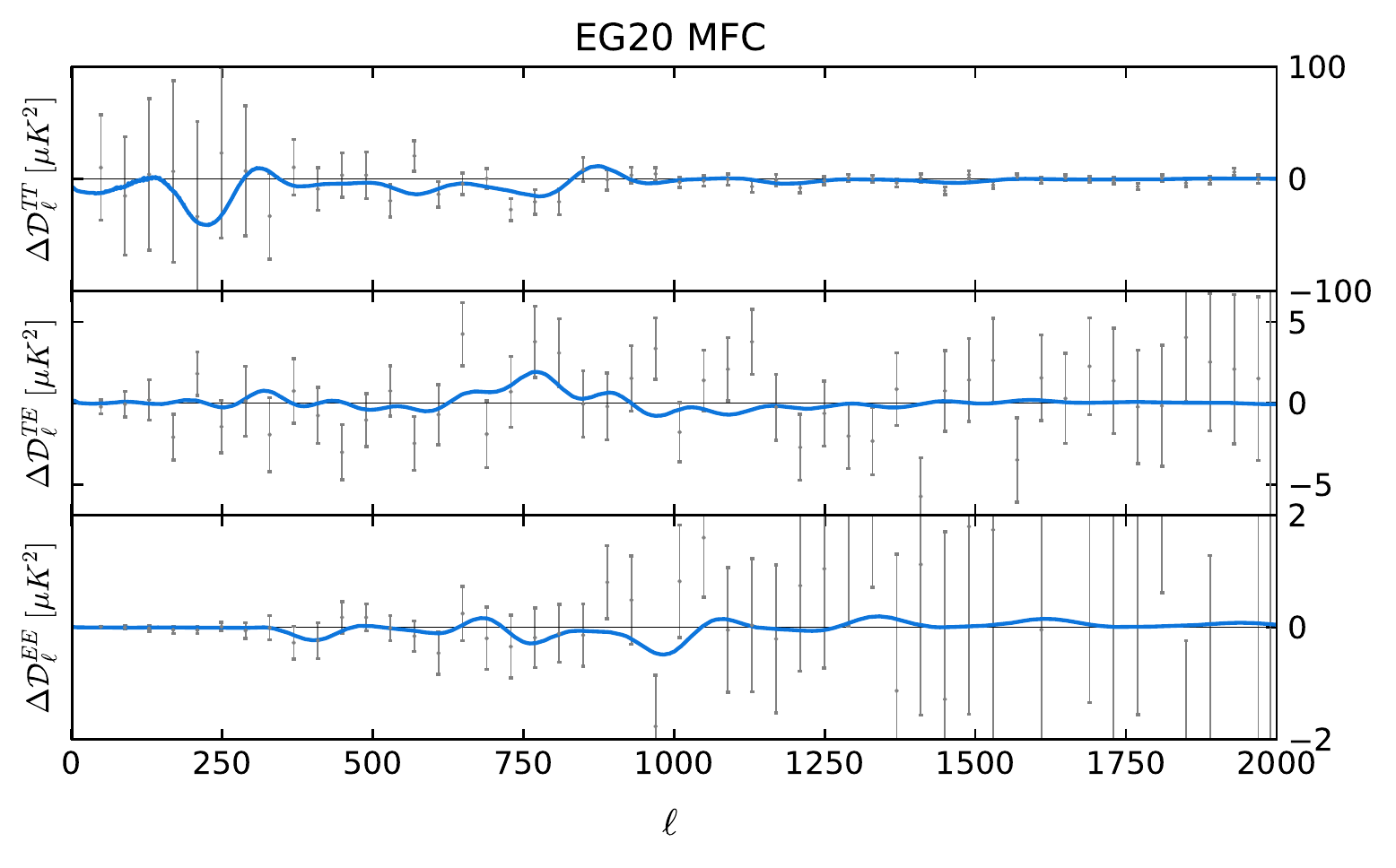}
\caption{\footnotesize\label{fig:Resid-EG20-MFC} Residual plots in the standard clock model for medium frequency TTTEEE best fit to EG20 likelihood. We find $\Delta\chi^2=7.5$ compared to baseline in method III.}
\end{figure}

\begin{figure}
\centering
\includegraphics[width=0.96\columnwidth]{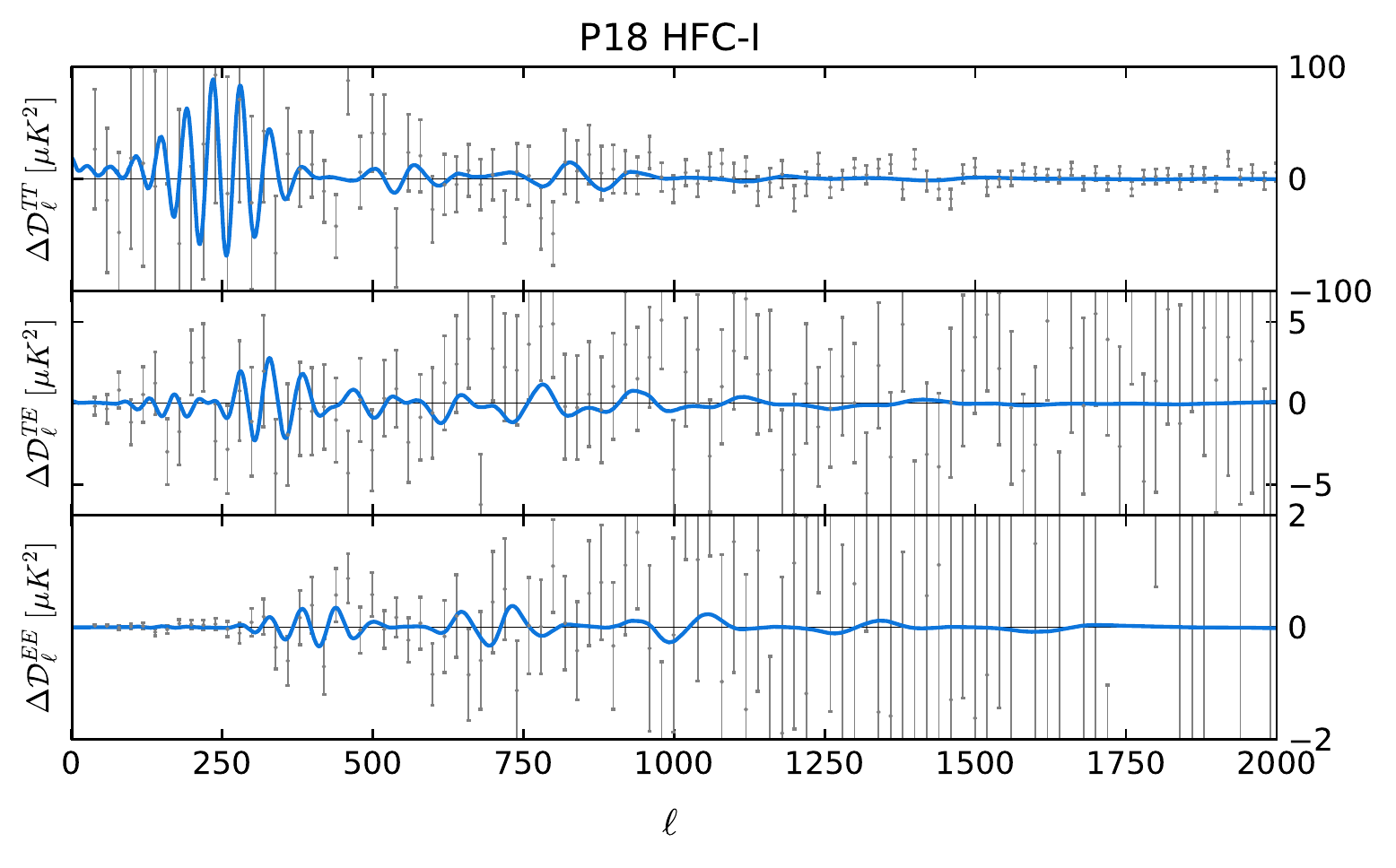}
\caption{\footnotesize\label{fig:Resid-P18-HFC1} Residual plot in the standard clock model for high frequency TTTEEE best fit candidate 1 to Plik bin1 likelihood. We find $\Delta\chi^2=8.8$ compared to baseline in method III.}
\end{figure}

\begin{figure}
\centering
\includegraphics[width=0.96\columnwidth]{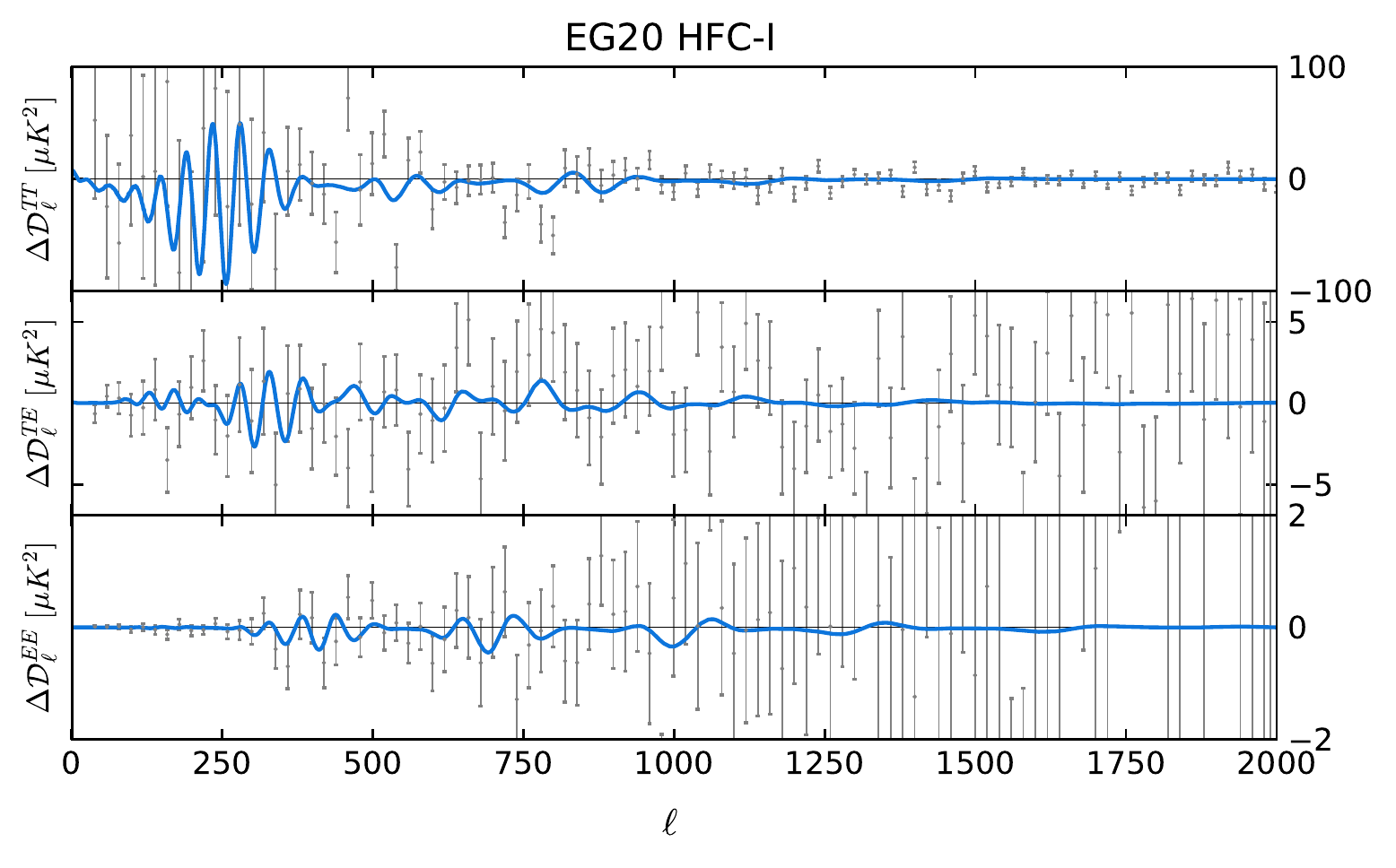}
\caption{\footnotesize\label{fig:Resid-EG20-HFC1} Residual plots in the standard clock model for high frequency TTTEEE best fit candidate 1 to EG20 likelihood. We find $\Delta\chi^2=8.3$ compared to baseline in method III.}
\end{figure}

\begin{figure}
\centering
\includegraphics[width=0.96\columnwidth]{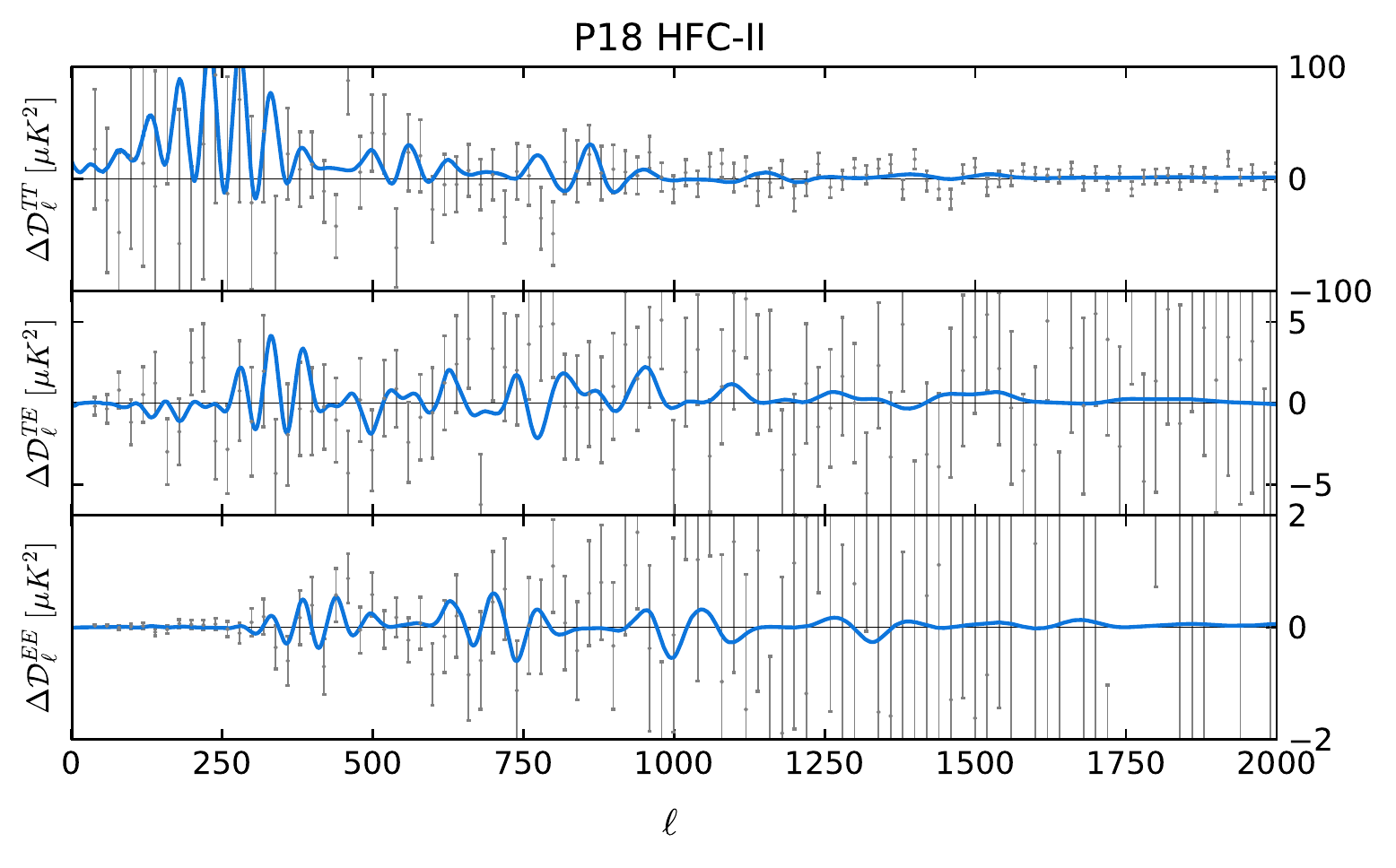}
\caption{\footnotesize\label{fig:Resid-P18-HFC2} Residual plot in the standard clock model for high frequency TTTEEE best fit candidate 2 to Plik bin1 likelihood. We find $\Delta\chi^2=15.7$ compared to baseline in method III.}
\end{figure}

\begin{figure}
\centering
\includegraphics[width=0.96\columnwidth]{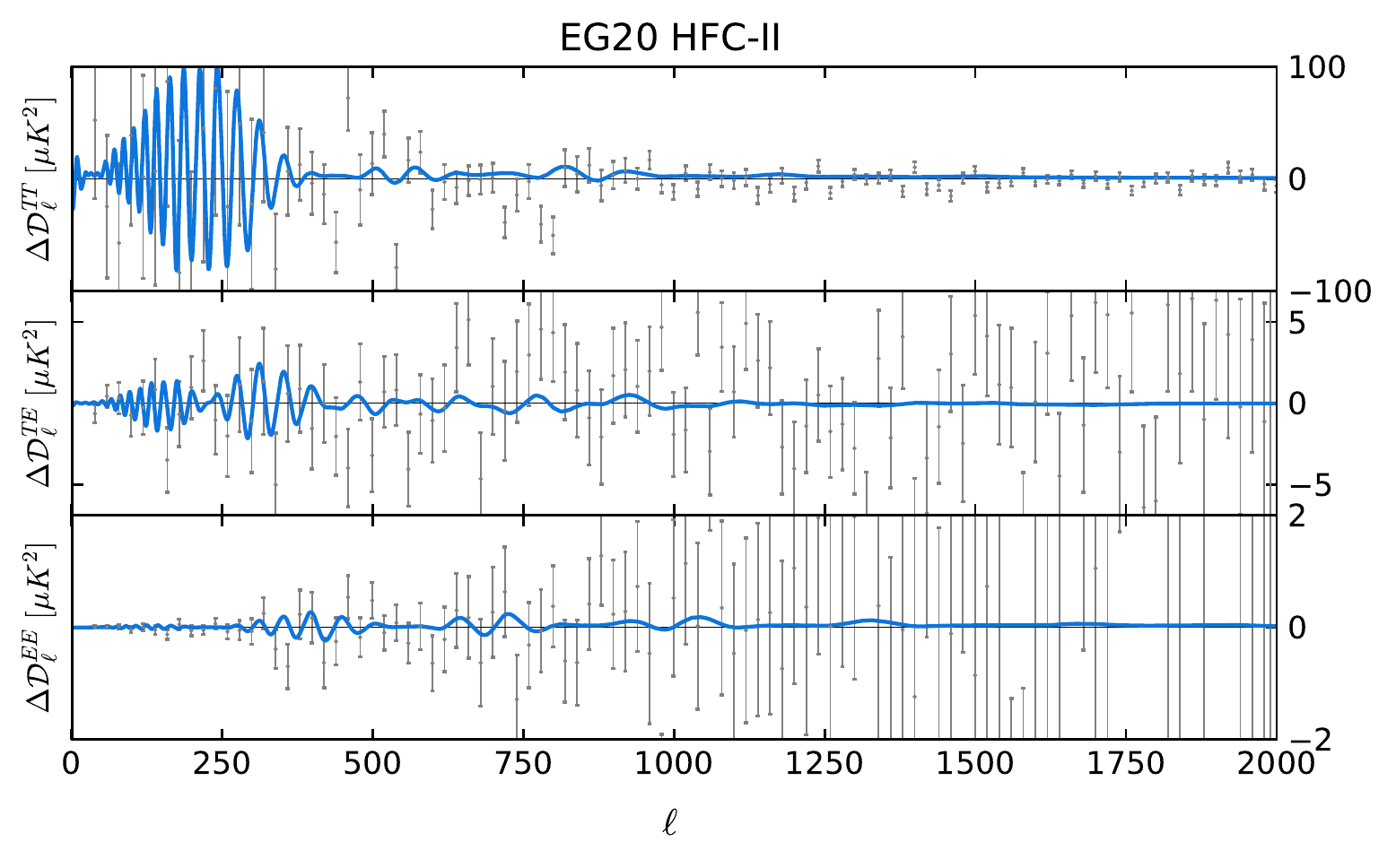}
\caption{\footnotesize\label{fig:Resid-EG20-HFC2} Residual plots in the standard clock model for high frequency TTTEEE best fit candidate 2 to EG20 likelihood. We find $\Delta\chi^2=8.5$ compared to baseline in method III.}
\end{figure}

\subsubsection{Mid frequency candidates}

\paragraph{TTTEEE candidates:}

The best fit candidate found in the medium frequency range occurs around $m_\sigma/H\sim 33$ in P18. In EG20 best fit shifts to $m_\sigma/H\sim 36$. For these two best fit spectra we plot the residual angular power spectra in Figs.~\ref{fig:Resid-P18-MFC} and \ref{fig:Resid-EG20-MFC} respectively. The residual binned data are now binned with $\Delta\ell=40$. The features mainly fit TT data around the first acoustic peak and around $\ell\sim700-1000$ multipoles better than baseline. The peak around $\ell=750$ in TE residual also seems to be addressed better by the MFC. Compared to LFC, the improvement in MFC is less, which is evident from the residual plots. The improvement in fit are very similar in P18 and EG20.

\paragraph{A comparison with TT candidates:}
In Fig.~\ref{fig:PPS-P18-All} the P18 TT+lowT+lowE best fit candidate is plotted in gold in the top-right panel. Similar to the LFC case, we find the features located at larger scales. At scales smaller than $0.045~\mathrm{Mpc^{-1}}$ the features match exactly with those of the MFC candidate to TTTEEE+lowT+lowE data. The matching makes it evident that improvement from MFC is mainly driven by TT data. Note that with similar frequency ($m_\sigma/H\approx 30$) but with higher amplitude, Ref.~\cite{Chen:2014joa,Chen:2014cwa} discussed the role of the CPSC model in addressing certain features in the multipole ranges $\ell=750-850$. While the MFC bestfit (see Figs.~\ref{PSC-6P-P18-deltachisquare} and~\ref{PSC-6P-EG20-deltachisquare}) is not exactly the same as the one in the aforementioned papers, they share some similarities basically because they are both fitting the similar features in the data. Besides the improvement in the data sets, this work is different from  \cite{Chen:2014joa,Chen:2014cwa} in the following ways. 
Firstly and most importantly, as we emphasized, the full numerical approach exploited in this work opens up significantly larger portion of parameter space available for data comparison, e.g.~allowing the variation in the clock field mass $m_\sigma/H$ along with other parameters. 
Secondly, the full numerical approach also demands that, in order to fit the realistic data, we have to use a quadratic potential instead of linear potential for the $\Theta$ field. The analytical template used in~\cite{Chen:2014cwa} was designed from a linear potential which exhibits some mismatch from the quadratic potential in the short scales. Because of these reasons we find the favored parameter space from Planck 2018 TT data near the MFC is shifted from $m_\sigma/H=30$ in earlier analyses along with other changes.

\subsubsection{High frequency candidates}
\paragraph{TTTEEE candidates:}

Since we compare the full CPSC model numerically, HFCs that are especially hard to capture with template approach, can now be tested against the data. The peaks in the $\Delta\chi^2$ sample plots, as well as the posterior distribution of $m_\sigma/H$, point towards a concentration of better likelihood around high frequency signals. From the samples, we have identified two HFC candidates that are referred to as HFC-I and HFC-II. In P18, HFC-I corresponds to $m_\sigma/H\sim53$. In Fig.~\ref{fig:Resid-P18-HFC1} we plot the residual angular power spectra. Residual data are plotted with a binwidth $\Delta\ell=20$. HFC candidates are interesting as they tend to fit sharp features in the data that are otherwise averaged out in the binned likelihood. Since we are using unbinned likelihood, the improvements are indeed obtained with fitting sharp oscillations and a smaller binwidth captures these features in the plot. In the TT residual we find that HFC candidates address `local' outliers that get averaged out with coarse binning. Similarly, for TE and EE, we note the features address the residual data at various scales. While HFC-I provides nearly 8 improvement in $\chi^2$, a sharper feature HFC-II with $m_\sigma/H\sim69$ represents the global best fit with nearly 16 improvement. Note that in method-II, we obtain 20 improvement in HFC-II, which again is not conservative as in both baseline and CPSC models we have neglected the $\chi^2$ from the priors. Compared to HFC-I, at intermediate and smaller scales ($\ell>500$), sharper oscillations in HFC-II help in additional improvement in fit to the data. Since sharp oscillation can also represent spurious features in the data coming from statistical fluctuations, more sensitive observations are likely to falsify these candidates.

Although presently we do not have a CMB observation more sensitive than Planck in the first few acoustic peaks, we do have the EG20 likelihood that is different from P18 and also claimed as the {\it most powerful} likelihood. The best fit candidates at high frequency from EG20 likelihood have masses around $m_\sigma/H\sim53$ (HFC-I) and $52$ (HFC-II). These two candidates are very different in amplitude and frequencies as can be seen by comparing the left and right  bottom panels in Fig.~\ref{fig:PPS-EG20-All}. EG20 HFC-I candidate is very similar to the P18 HFC-I candidate, as also shown by the residual plot in Fig.~\ref{fig:Resid-EG20-HFC1}. The improvement in fit is also similar. EG20 HFC-II candidate has the oscillations located at larger scales and therefore the main improvement in fit comes from the feature at the first acoustic peak as plotted in Fig.~\ref{fig:Resid-EG20-HFC2} TT residual. In TE residuals we find the HFC-II candidate to address residual data at $\ell<200$ with sharp oscillations and around $\ell\sim300-400$ with relatively broader oscillations. This candidate also attempts to fit EE data around $\ell\sim400$. From the PPS, a suppression of power at large scales $k<10^{-3}\,\mathrm{Mpc}^{-1}$ can be noticed, fitting the low-$\ell$ TT angular power spectrum marginally better than the baseline. Note that the distinct locations of these two candidates with very similar masses can  be clearly noticed in the sample plots in TTTEEE panel of Fig.~\ref{PSC-6P-EG20-deltachisquare}. Since P18 HFC-II represents the global best fit to the Plik likelihood, we also searched for the best fit starting from that best fit in EG20. However we find that the candidate no longer fits EG20 data better. This is a crucial comparison demonstrating a particular best fit candidate not agreeing to another likelihood from the same observation. This mismatch indicates that the P18 HFC-II  is more likely a spurious feature than the P18 HFC-I.

\paragraph{A comparison with TT candidates:}
The high frequency candidates for TT are also different compared to the TTTEEE candidates. For a similar $m_\sigma/H$, the TT candidate at high frequency ($m_\sigma/H\sim52$) prefers the features at larger scales with higher amplitude. The candidate is plotted in gold in the bottom left panel of Fig.~\ref{fig:PPS-P18-All}. A large $\xi\sigma_f$, in addition to enhancing the amplitude of the oscillations, also suppresses the power spectra at large scale which helps improve the fit to lowT data. Since the large amplitude oscillation, which also extends to small scales, does not agree with TE and EE data, we do not find this candidate as the best fit to P18 TTTEEE+lowT+lowE data. We also identify another best fit close to P18 HFC-II in the TT likelihood, but at reduced frequency ($m_\sigma/H\sim63$). Here too the best fit candidate has  oscillations shifted to larger scales compared to P18 HFC-II (see bottom right panel of Fig.~\ref{fig:PPS-P18-All}).

\subsubsection{Comparison of feature candidates between Plik bin1 (Planck 2018) and Clean CAMSpec (EG20) likelihoods}
Given the power spectra of temperature and polarization anisotropies and their correlation, we have the opportunity of cross examining a best fit candidate to one spectra with others. Features that are supported by all the datasets have the best chance to emerge as {\it true feature}. Our analysis with different combination of datasets in both P18 and EG20 likelihoods helps us to pick particular candidates that we have discussed. Since these likelihoods are different in nature, a common feature in both likelihoods increases its significance. Comparing four best fit candidates at 3 different frequency regimes, we find that the LFC, MFC and HFC-I provides similar improvement in fit to both likelihoods. HFC-II candidate, that represents the global best fit to the P18 likelihood, is no longer supported in the EG20 likelihood. Since LFC candidate provides $11.4$ and $9.7$ improvement in $\chi^2$s compared to the baseline best fit in P18 and EG20 likelihood to the complete TTTEEE+lowT+lowE data respectively, we can claim this as the global best fit of this particular CPSC model to the final release of Planck data. 

Note that, while we discussed only four best fit candidates, we should clarify that given the plethora of feature candidates the CPSC model can generate, it is also possible to find other candidates with different frequencies and amplitudes of the oscillations that occur at different scales. We have presented the spectra that are qualitatively different from each other and give substantial improvement in fit to the data. As can be noticed from the TE and EE panels of Figs.~\ref{PSC-6P-P18-deltachisquare},~\ref{PSC-6P-P18-heatmap},~\ref{PSC-6P-EG20-deltachisquare},~\ref{PSC-6P-EG20-heatmap} there are several local best fits to the individual data coming from different regions in the parameter space. However, since the signal-to-noise ratios of these individual data (mainly EE) are lower than TT and therefore the complete TTTEEE data, these feature candidates are unlikely to stand out unless they overlap with TT best fits.

\section{Conclusion and discussions}
\label{Sec:Conclusion}

The final release of Planck data provides the most precise CMB temperature and polarization anisotropy power spectra and their correlations ranging from the largest scales to small scales obtained from a single survey, with cosmic variance limited temperature data. Since the primordial power spectra can be best constrained by CMB surveys, undoubtedly before the next full sky CMB mission from LiteBIRD~\footnote{\href{http://litebird.jp/eng/}{http://litebird.jp/eng/}}, Planck constraints on primordial features will remain the most relevant constraints over the largest range of cosmological scales. With the new datasets and two different likelihoods released during and after Planck 2018 final release, namely Plik bin1 likelihood and clean CamSpec likelihood (referred to as P18 and EG20 respectively in the main text), the main purpose of this paper is to use a specific inflationary model of CPSC as an example to test the methodology of directly and efficiently compare the numerical full model prediction with data.

The inflationary CPSC signature in the primordial power spectrum consists in a sharp feature signal with a characteristic sinusoidal running in momentum space followed by a clock signal with an inflationary resonant running on smaller scales. The properties of such a signal, however, depend in a complicated fashion on the parameters of the model, and the full comparison with data has remained as an open challenge. In this work, we have successfully implemented a methodology that starts out with numerical predictions of the full model, bridged by effective parameters constructed out of physical ones appearing in the model Lagrangian, and finishing off with data comparison by the method of nested sampling in different oscillation frequency bands. The method provides an efficient exploration of the large prior volume and allows us to constrain the amplitude, the location and, most importantly, the frequency of the oscillations of the clock signal. 
  
\begin{figure}
\centering
\includegraphics[width=0.96\columnwidth]{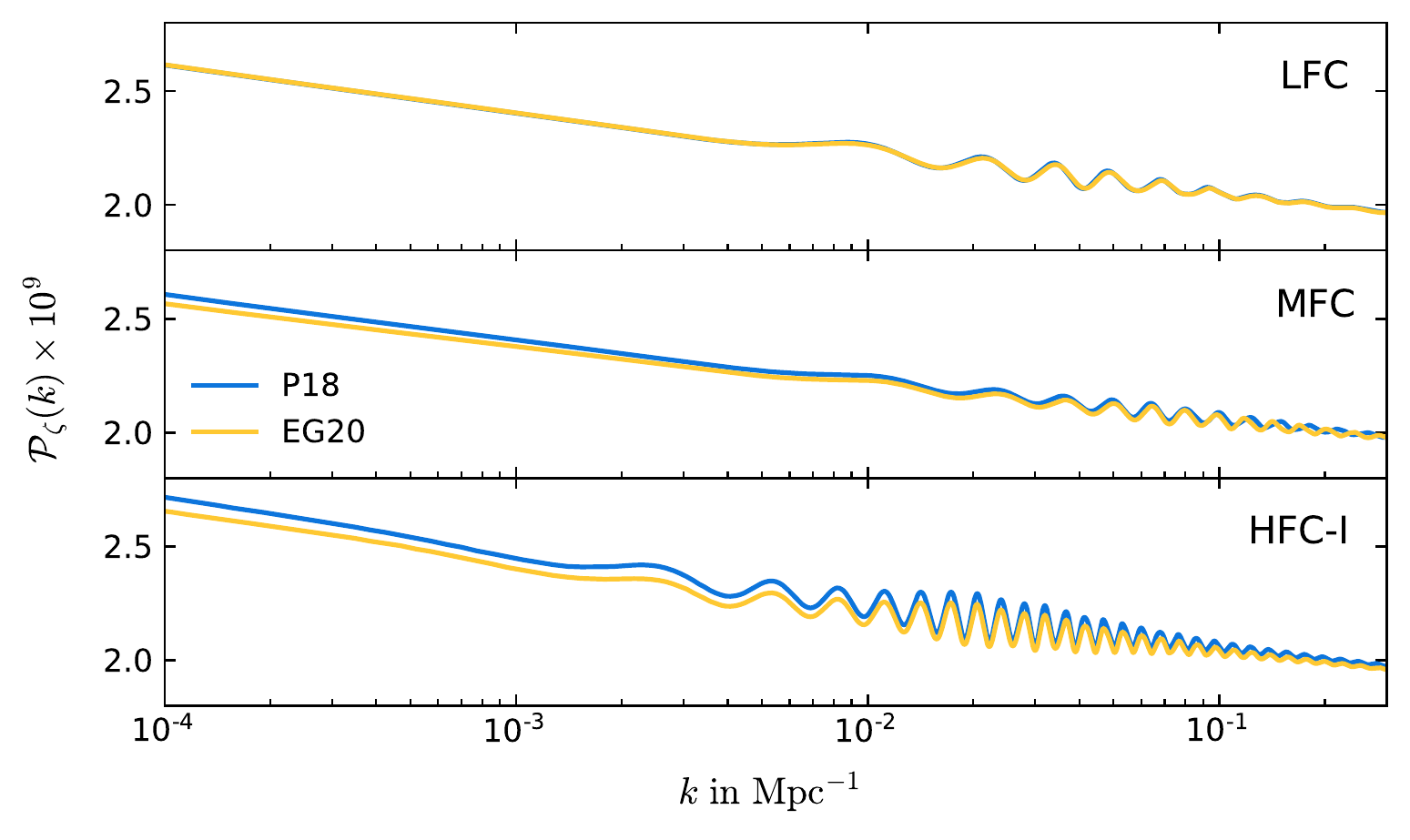}
\caption{\footnotesize\label{fig:bestfitcomparison} Comparison of the LFC, MFC and HCF-I for P18 and EG20.}
\end{figure}

Our analysis shows multimodal posterior distributions for the effective parameters that allow the model to fit the data better than power law spectrum and to address different data residuals depending on the frequency of the signal or, in other words, on the effective mass of the clock field. Combining all the temperature and polarization datasets and comparisons between P18 and EG20 provides powerful constraints that can eliminate many candidates from individual dataset such as TT.  We demonstrate that at least three best fit candidates, supported by both P18 and EG20, emerge from our analysis. Fig.~\ref{fig:bestfitcomparison}, where we plot these best fit candidates, shows that the candidate spectra look very similar for P18 and EG20. In particular, the low frequency one is exactly the same  for P18 and EG20 and provides the global best fit to both data sets with an improvement in $\Delta\chi^2\sim10$ with respect to the baseline featureless model. Nonetheless, the difference of all three candidates in $\chi^2$ is small and there are other candidates that have similar $\chi^2$ that we did not highlight. This is an indication that these feature candidates are still consistent with statistical fluctuations.
Consistently, despite the improvement in the fits, our model selection analysis shows a small, but negative, Bayes factor, which takes into account look-elsewhere effect by allowing maximum prior volume for the features determined by theoretical limits and observational windows. We conclude that no evidence for features of this CPSC model is found in Planck data.

The low statistical significance of these candidates is also consistent with the fact that these anomalies in the Planck residual data at short scales have been fit by other (single field) primordial features models such as resonant models (ripples in potential)~\cite{Akrami:2018odb}, pure sharp feature models~\cite{Hazra:2014goa,Fergusson:2014hya,Fergusson:2014tza,Domenech:2019cyh,Canas-Herrera:2020mme}, or features from alternative-to-inflation models \cite{Domenech:2020qay}. Future data will determine if any of these candidates will survive or distinguish the differences between them.

The method used in this paper can be applied to other complicated feature models where the analytical prediction is unavailable, including both inflationary and non-inflationary models, to fully exploit the exquisite constraining power of Planck. For example, while the clock signal is expected to depend only very weakly on model details, changing the mechanism that excites the massive field oscillations can significantly alter the envelop of the sharp feature signal and the relative relation between the clock signal and the sharp feature signal. The methodology of this paper makes the data comparison of these model variations possible.  
 An intriguing observation that further motivates this exploration is that the location of the sharp feature of the three candidates are all around $k_0\sim0.5 - 2 \times 10^{-3}$ Mpc$^{-1}$, see Tab.~\ref{tab:candidates}. This is close to the location of the known sharp feature candidate around $\ell \sim 20$ \cite{Peiris:2003ff}. However, the oscillation amplitudes of our candidates around this scale are all too small to provide any noticeable fit to this feature. This suggests that a different model building may provide a better overall fit. Work in this direction is in progress \cite{CPSC_Dip}. 

The cosmological scales of most of the best fit candidate features in our analysis have substantial overlap with those probed by small scale CMB observations, such as polarization observation from ACT~\cite{Aiola:2020azj}, and with LSS observations with upcoming Euclid and LSST.  While Planck data is cosmic variance limited in temperature, the polarization data however is not. With upcoming LiteBIRD and proposed CMBBHARAT~\footnote{\href{http://cmb-bharat.in/}{http://cmb-bharat.in/}}, it is possible to probe different features with much higher statistical significance.
Feature models with features extending to intermediate and small scales can be constrained better if LSS surveys are analyzed jointly with CMB data \cite{Chen:2016vvw,Ballardini:2016hpi,Palma:2017wxu,LHuillier:2017lgm,Ballardini:2017qwq,Beutler:2019ojk,Ballardini:2019tuc,Debono:2020emh,Li:2021jvz}.
We plan to explore the probabilities with which different feature candidates can be identified or ruled out using upcoming CMB and LSS observations. 

Apart from the two-point statistics, bispectra in feature models are expected to exhibit features that have highly correlated scale-dependence with that in the power spectrum \cite{Chen:2010xka}. Therefore, if the sizes of the bispectra from these models are large enough, a joint analysis of the power spectrum and bispectrum can significantly improve the statistical significance \cite{Fergusson:2014hya,Fergusson:2014tza}.

Finally, an alternative approach to one of the steps of our methodology that may be worth trying is to replace the direct numerical computation of the model prediction with a pre-computed template bank which covers all interesting parameter space with fine grids. This approach may be particularly useful if the numerical step in the pipeline becomes computationally too costly.

	\section*{Acknowledgements}

We thank Steven Gratton and Will Handley for important discussions related to CamSpec and Polychord. Most of the computations in this paper were run on the FASRC Cannon cluster supported by the FAS Division of Science Research Computing Group at Harvard University. Some computations were also run in Institute of Mathematical Science’s High Performance Computing facility (hpc.imsc.res.in) [Nandadevi].

\appendix

	\bibliographystyle{JHEP}
	\bibliography{SCPlanck}

\end{document}